\documentclass[aps,prev,twocolumn,preprintnumbers,floatfix,nofootinbib]{revtex4-1}

\usepackage{graphicx}
\usepackage{bm}
\usepackage{times}
\usepackage{slashed}
\usepackage{color}
\usepackage{epsfig}\usepackage{dcolumn}
\usepackage{color}
\def\br{\begin{eqnarray}}
\def\er{\end{eqnarray}}
\def\be{\begin{equation}}
\def\ee{\end{equation}}

\newcommand{\zp}{Z^{\prime}}

\begin{document}

\preprint{LPT-Orsay-18-72}

\title{XENON1T Takes a Razor to a Dark $E_6$-Inspired Model}

\author{Daniel A. Camargo$^1$}
\author{Yann Mambrini$^{2}$}
\author{Farinaldo S. Queiroz$^{1}$}

\email{farinaldo.queiroz@iip.ufrn.br}

\affiliation{$^1$International Institute of Physics, Universidade Federal do Rio Grande do Norte, Campus Universit\'ario, Lagoa Nova, Natal-RN 59078-970, Brazil\\
$^2$Laboratoire de Physique Th\'eorique (UMR8627), CNRS, Univ. Paris-Sud, Universit\'e Paris-Saclay, 91405 Orsay, France
}

\begin{abstract}
Motivated by the recent result of XENON1T collaboration with full exposure, 279 life days, that sets the most stringent limit on the spin-independent dark matter-nucleon scattering cross section we discuss a dark $E_6$-inspired model that features the presence of a $U(1)_{d-u}$ gauge symmetry. The dark matter candidate is a Dirac fermion that interacts with Standard Model fermions via a massive $Z^\prime$ that preserves the quantum number assignments of this symmetry. We compute the spin-independent scattering cross section off xenon nucleus and compare with the XENON1T limit; find the LHC bound on the $Z^\prime$ mass as well as the projection sensitivity of high-energy and luminosity LHC; and derive the Fermi-LAT bounds on the dark matter annihilation cross section based on the observation of gamma-rays in the direction of Dwarf Spheroidal galaxies. We exploit the complementarity between these datasets to conclude that the new bound from XENON1T severely constrain the model, which combined with the LHC upgrade sensitivity rules out this WIMP realization setup below $5$~TeV.
\end{abstract} 
    
\maketitle

\section{Introduction}
\label{introduction}

The presence of dark matter (DM) has been established by numerous observations stemming from different scales and epochs in the history of the universe \cite{Bertone:2004pz}. Apart from gravitation effects, we have not been able to detect the interactions involving dark matter particles and for this reason its nature remains unknown \cite{Queiroz:2016sxf,Kavanagh:2017hcl}. There are several dark matter candidates in the literature with masses spanning over several orders of magnitude \cite{Bergstrom:2009ib,Klasen:2015uma}. However, massive particles that are thermally in the universe are theoretically well motivated the most popular (massive particles that are thermally produced stays the most popular)\cite{Jungman:1995df,Bergstrom:2009ib,Baer:2014eja}. They are commonly referred to as WIMPs (Weakly Interacting Massive Particles) \cite{Arcadi:2017kky}.\\

WIMPs are subject of intense searches that involved different techniques and observables which are classified as direct, indirect and collider probes \cite{Cushman:2013zza,Queiroz:2016awc}. Direct detection refers to the measurement of the recoil energy deposited in nuclear scatterings by WIMP-like particles on their scattering off of nuclei target used in underground experiments \cite{Undagoitia:2015gya}. Using discriminating variable to distinguish nuclear recoil from electron recoil and a good control over background from various sources, direct detection experiments become discover machines since the scattering rate depends on the properties of dark matter particle, such as the scattering cross section and mass \cite{Gondolo:1996qw,Pato:2010zk,Freese:2012xd,DelNobile:2014sja}, and the local dark matter density \cite{Catena:2009mf,Pato:2010yq,Iocco:2011jz} . Therefore, null results from direct detection experiments can be interpreted as upper limits on the pair WIMP-nucleon scattering cross section and dark matter mass \cite{Kelso:2011gd}. We have several experiments searching for WIMP events using different readout techniques and targets that make them complementary to each other and sensitive to different regions of the scattering cross section {\it vs} dark matter mass parameter space  \cite{Akerib:2013tjd,Agnese:2013jaa}.\\

Indirect detection covers to the observation of gamma-rays or other possible final states resulted from DM annihilation or decay \cite{Scott:2009jn,Abdo:2010ex,Hooper:2012sr,Gonzalez-Morales:2014eaa,Mambrini:2015sia,Baring:2015sza,Li:2015kag,Drlica-Wagner:2015xua,Zhao:2016xie,Garcia-Cely:2016pse,Li:2018kgy}. Fermi-LAT has been continuously observing the gamma-ray emission in the direction of over a dozen of Dwarf Spheroidal galaxies which are dominated by dark matter, and since no significant excess has been observed, restrictive limits have been placed on the gamma-ray flux resulted from dark matter annihilations\cite{Ackermann:2015zua,Ackermann:2015zua,Ahnen:2016qkx}. The gamma-ray flux originated from dark matter annihilation depends on the dark matter density distribution known as density profile; the gamma-ray energy spectrum that accounts for how many gamma-rays are produced per annihilation; on the annihilation cross section and dark matter mass \cite{Profumo:2013yn,Lopez:2015uma,DeAngelis:2017gra,Acharya:2017ttl}. The gamma-ray flux produced by dark matter annihilation is very important especially for dark matter models that rely entirely on the thermal production of dark matter and the dark matter annihilation cross section is velocity independent as it happens to be the case for our dark $E_6$-inspired model \cite{Slatyer:2015kla,Slatyer:2016qyl,Liu:2016cnk,Bernal:2016guq,Dutra:2018gmv}.\\

Collider searches for WIMPs are characterized by events featuring large missing energy accompanied by visible counter-parts \cite{Goodman:2010ku,Rajaraman:2011wf,Bauer:2013ihz,Proceedings:2016rni}. Sometimes the most efficient method to probe a dark sector is not by searching for the dark matter itself at colliders, especially when it has mediators involved have sizable couplings to SM fermions. In the model we consider, the massive $Z^\prime$ does have sizable couplings to leptons and thus it is subject to stringent bounds stemming from the LHC \cite{Aad:2014cka,Khachatryan:2016zqb,Sirunyan:2018exx}. \\

With these dark matter searches in mind, a multitude of phenomenological studies have been carried out to illustrate the viable parameter based on simplified dark matter models without specifying a gauge group.  One of the most attractive is the Dirac fermion charged under an $U(1)$ gauge symmetry. In this context, the $Z^{\prime}$ gauge boson is a natural mediator between the dark and the visible sector. \\

We advocate that it is more interesting to fix a gauge setup which is motivated by theoretical constructions and assess whether a viable dark matter candidate is possible without worrying about the further complications arising from a UV completion \cite{Lopez:1991pp,Baer:2008jn,Baer:2008eq,Agashe:2009ja,Chakrabortty:2013voa,Anandakrishnan:2013tqa,Ko:2015fxa,Arbelaez:2015ila,Chen:2017rpn,Arcadi:2017atc,Coriano:2017ghp,King:2017cwv}. That said, we choose to explore dark matter complementarity in the context of collider, direct and indirect detection searches in a specific setup where the dark matter particle is a Dirac fermion that interacts with a $Z^\prime$ boson that possesses interactions with the SM particle as dictated by the $U(1)_{d-u} $ gauge structure which appears in grand unification studies of the $E_6$ group \cite{Patrignani:2016xqp}.\\

The paper is organized as follows: In section I we introduce the model; in section II we discuss the dark matter observables; in section III we exhibit our findings before concluding in section IV.

\section{Model}
\label{sec1}

The origin of a $Z^\prime$ gauge boson is behind a new $U(1)$ gauge symmetry. The charges of the SM fermions under the $U(1)$ gauge symmetry are in principle arbitrary but should respect $SU(2)$ invariance with the left-handed up and down quarks having the same quantum number \cite{Bell:2016ekl,Bell:2016obu}. The same logic applied to the left-handed charged leptons and their corresponding neutrino flavors. Since the new interaction is not asymptotically
free in a UV completion, the model should be well-behaved at high energies and thus anomaly-free \cite{Ismail:2016tod,Ismail:2017ulg}. The anomaly cancellation can be highly non-trivial and for this reason  it is theoretically interesting to consider gauge groups that are generation-independent such as the $U(1)_{d-u}$ group that arises after spontaneous symmetry breaking in some $E_6$ constructions \cite{Patrignani:2016xqp}. The $E_6$ symmetry is theoretically motivated because it offers a hospital environment for unification of forces and incorporates interesting features of $SU(5)$ or $SO(10)$ groups concerning the generation of fermion masses and flavor symmetries \cite{Cho:1998nr,Nie:2001ti,Howl:2007zi,Stech:2008wd,Rojas:2015tqa,Benli:2017eld,Dutta:2018qei}, super heavy dark matter \cite{Schwichtenberg:2017xhv}, and collider physics \cite{Rojas:2015tqa}.\\

In order to incorporate dark matter we introduce a Dirac particle we write down the simplified Lagrangian that describes the dark matter interactions with the  $\zp$ and SM fermion as,

\begin{equation}
\label{equation:main}
\mathcal{L} \supset Z^\prime_\mu~\big[ \bar{\chi} \gamma^\mu \left( g_{\chi \text{v}} + g_{\chi a} \gamma^5 \right) \chi +  \sum_{f \in \text{SM} } \bar{f} \gamma^\mu \left( g_{f \text{v}} + g_{f a} \gamma^5 \right) f \big]
~,
\end{equation}where the sum is over all quarks and leptons. The $\zp$-SM couplings in Eq.\ref{equation:main} are determined by the SM fermion charges under the new gauge symmetry as follows, 

\begin{eqnarray}
\label{equation: SMcouplings}
g_{u \text{v}} &= \frac{g_{\zp}}{2} \left( z_{u_R} + z_{Q_L} \right) ~,~  g_{u a} = \frac{g_{\zp}}{2}  \left( z_{u_R} - z_{Q_L} \right)
\nonumber \\
g_{d \text{v}} &= \frac{g_{\zp}}{2} \left( z_{d_R} + z_{Q_L} \right) ~,~  g_{d a} = \frac{g_{\zp}}{2}  \left( z_{d_R} - z_{Q_L} \right)
\nonumber \\
g_{l \text{v}} &= \frac{g_{\zp}}{2} \left( z_{e_R} + z_{l_L} \right) ~~,~~ g_{l a} = \frac{g_{\zp}}{2} \left( z_{e_R} - z_{l_L} \right)
\nonumber \\
g_{\nu \text{v}} &= \frac{g_{\zp}}{2}   z_{l_L} ~~~~~~~~~~~~~,~~  g_{\nu a} = -\frac{g_{\zp}}{2} z_{l_L}
~.
\end{eqnarray}

The $z$'s charges of the fields as shown in Table I.  In our model (Eq.\ref{equation:main}) we are assuming that the dark matter candidate is a chiral field. Therefore the gauge anomaly cancellation becomes non-trivial and the mechanism behind the anomaly cancellation and $Z^\prime$ mass can change our numerical findings \cite{Kahlhoefer:2015bea,Celis:2016ayl,Duerr:2016tmh,Jacques:2016dqz,Bell:2016ekl,Bell:2016uhg,DEramo:2016gos,Englert:2016joy,Brennan:2016xjh,DeRomeri:2017oxa,Ismail:2017ulg,Cui:2017juz,Mantilla:2017ijh,Bauer:2018egk}. We will assume that Eq.\ref{equation:main} offers a good description of the dark matter phenomenology and to simplify our analysis we adopt $g_{\chi v}=g_{\chi a}$. The scenarios with $g_{\chi v}=g_{\chi a}$ and the $g_{\chi a}=0$ share a similar dark matter phenomenology and this setup with $g_{\chi a}=0$ is equivalent to the one of vector-like dark matter fermion. In other words, the dark matter observables and collider bounds derived here are similar to the one present in vector-like fermion dark matter setups. Moreover, we adopt $g_{\zp}=1$ throughout. That said, Eq.\ref{equation:main} will dictate the entire phenomenology of our work as we describe hereafter.

We point out that $E_6$ group breaks into  $SU(5) \otimes U(1)\otimes U(1)$. Depending on how the spontaneous symmetry is broken you can generate different $U(1)$ models. The most common breaking  $E_6 \rightarrow SU(5) \otimes U(1)_{\chi} \otimes U(1)_{\eta}$, as adopted by the CMS and ATLAS collaborations, where the physical $Z^{\prime}$ is given by $Z^\prime = Z^\prime_\psi \cos\theta + Z^\prime_\chi\sin\theta$, where $\theta$ is the mixing angle which can be in principle suppressed. Our model $U(1)_{d-u}$ resembles the $Z^\prime$ boson of the $U(1)_\psi$ model because the magnitude of $Z^\prime$ couplings to right-handed fermions is the same, but the $U(1)_{d-u}$ model does not have couplings to lepton-handed fields, and for this reason the $U(1)_{d-u}$ model is also known as the right-handed $U(1)$. Anyways, the dark matter phenomenology is dominated by the $Z^\prime$ neutral current, thus we do not expect significant differences between our model and the $U(1)_\psi$ model. 

\begin{table}[h]
\centering
\begin{tabular}{| c| c | c | c | c |}
\hline
$Q_L$ & $u_R$ & $d_R$ & $l_L$ & $e_R$\\
$0$ & $-1/3$ & $1/3$ & $0$ & $1/3$\\
\hline
\end{tabular}
\caption{Standard Model fermions $U(1)_{d-u}$ charges.}
\label{table: charge}
\end{table}

\subsection{Direct Detection}
\label{section: directdetection}

The Lagrangian with a dark matter field coupling chirally via a $Z^\prime$ such as the one presented in Eq.\ref{equation:main} leads to a dark matter-nucleus scattering that is both spin-independent and spin-dependent. The spin-independent process is coherent,  so it scales with the atomic mass of the nucleus. On the other hand, the spin-dependent scattering goes with the net spin of the nucleus which is not much far from one. Therefore, bounds on spin-independent scattering are much more stringent for heavy nuclei such as xenon that has an atomic mass $A=131$. For this reason, we will concentrate our direct detection discussion on spin-independent scattering. The WIMP-nucleon spin-independent scattering cross section in this case is, 

\begin{equation}
\sigma^{SI}_n = \frac{\mu^2_{\chi n}}{\pi} \left( \frac{Z f_{p} + (A-Z) f_{n}}{A} \right)^2, 
\end{equation}where,

\begin{equation}
f_{p} =\equiv \frac{g_{\chi v}}{m_{\zp}^2} \left( 2 g_{u v} + g_{ d v} \right)
\end{equation}and

\begin{eqnarray}
f_{n} &\equiv \frac{g_{\chi v}}{m_{\zp}^2} \left( g_{u v} + 2 g_{d v} \right),
\end{eqnarray}with $\mu_{\chi n}$ being the WIMP-nucleon reduced mass, $Z$ the atomic number of the target nucleus. \\

With the theoretical prediction at hand, we can compare our finding with the current limit from XENON1T collaboration presented at MPIK on May 28, 2018 \footnote{\url{https://indico.cern.ch/event/726877/}} which was later followed by the XENON1T publication \cite{Aprile:2018dbl}. The experiment limit is based on 278.8 life days exposure with 1.3 tons fiducial mass. The expected background events are 0.75 neutrons,  5.36 surface events, 0.02 from neutrino-nucleus coherent background among other that total about $8.4\pm 0.63$ events. The collaboration observed about 11 events in the signal region thus with a small excess of events. Because of the presence of such excess of events the scientific goal of the collaboration of achieving a sensitivity to the spin-independent scattering cross section down to $1\times 10^{-47} \mathrm{cm^{2}}$ was not achieved but the collaboration successfully carried their analysis and set the world leading upper limit on the WIMP-nucleon spin-independent scattering cross section of $4\times 10^{-47} \mathrm{cm^2}$ \cite{Aprile:2018dbl}. These results severely constrain our model as shown in figures 1-3. The blue hashed region is represent the XENON1T exclusion region \footnote{There are competitive limits from other experiments \cite{Akerib:2013tjd,Akerib:2016vxi,Tan:2016zwf,Cui:2017nnn}.}. 

\begin{figure}
\includegraphics[width=\columnwidth]{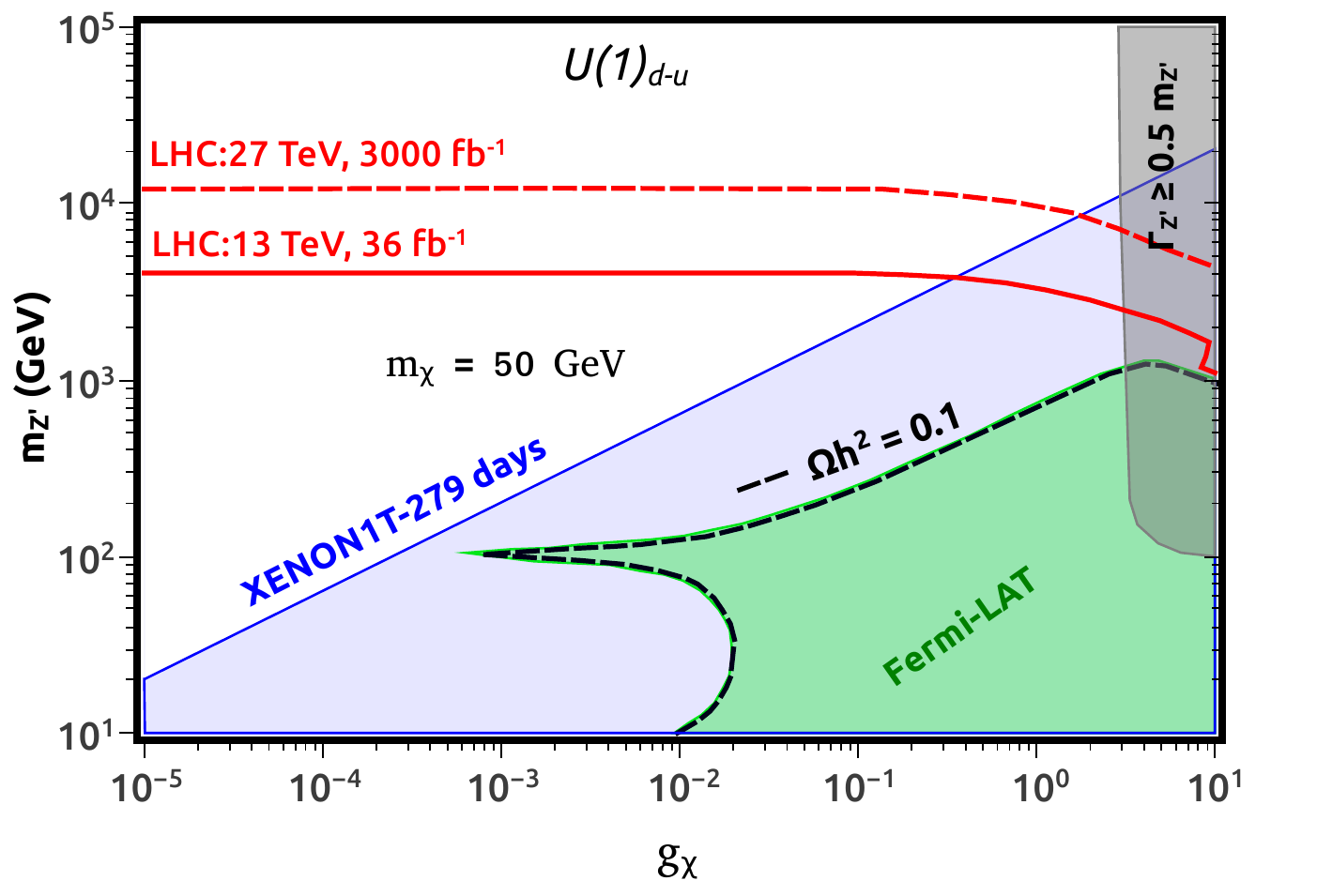}
\caption{Dark matter complementarity plot for $m_{\chi}=50$~GeV. In blue we show the excluded region based on the recent XENON1T bound on the spin-independent scattering cross section. In green, the region ruled out by Fermi-LAT through the observation of dwarf spheroidal galaxies. The red horizontal curve is the LHC limit on the $Z^\prime$ mass using dilepton data at $13$~TeV. The black dashed curve delimits the region of parameter space that yields the correct relic density.
%via the freeze-out mechanism
}
\label{fig1}
\end{figure}

\begin{figure}
\includegraphics[width=\columnwidth]{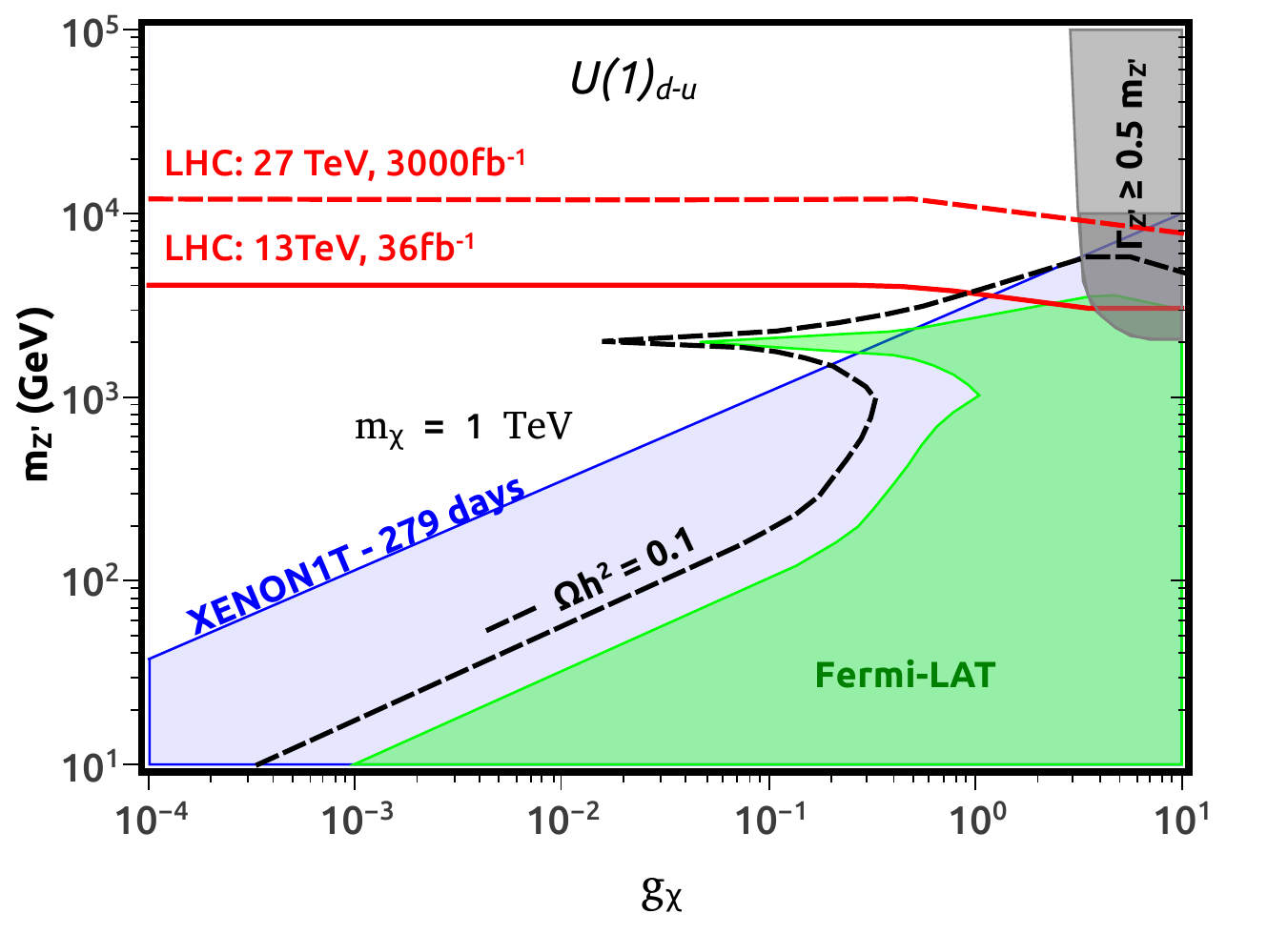}
\caption{Same than figure \ref{fig1} but for $m_{\chi}=1$~TeV.}
\label{fig2}
\end{figure}

\begin{figure}
\includegraphics[width=\columnwidth]{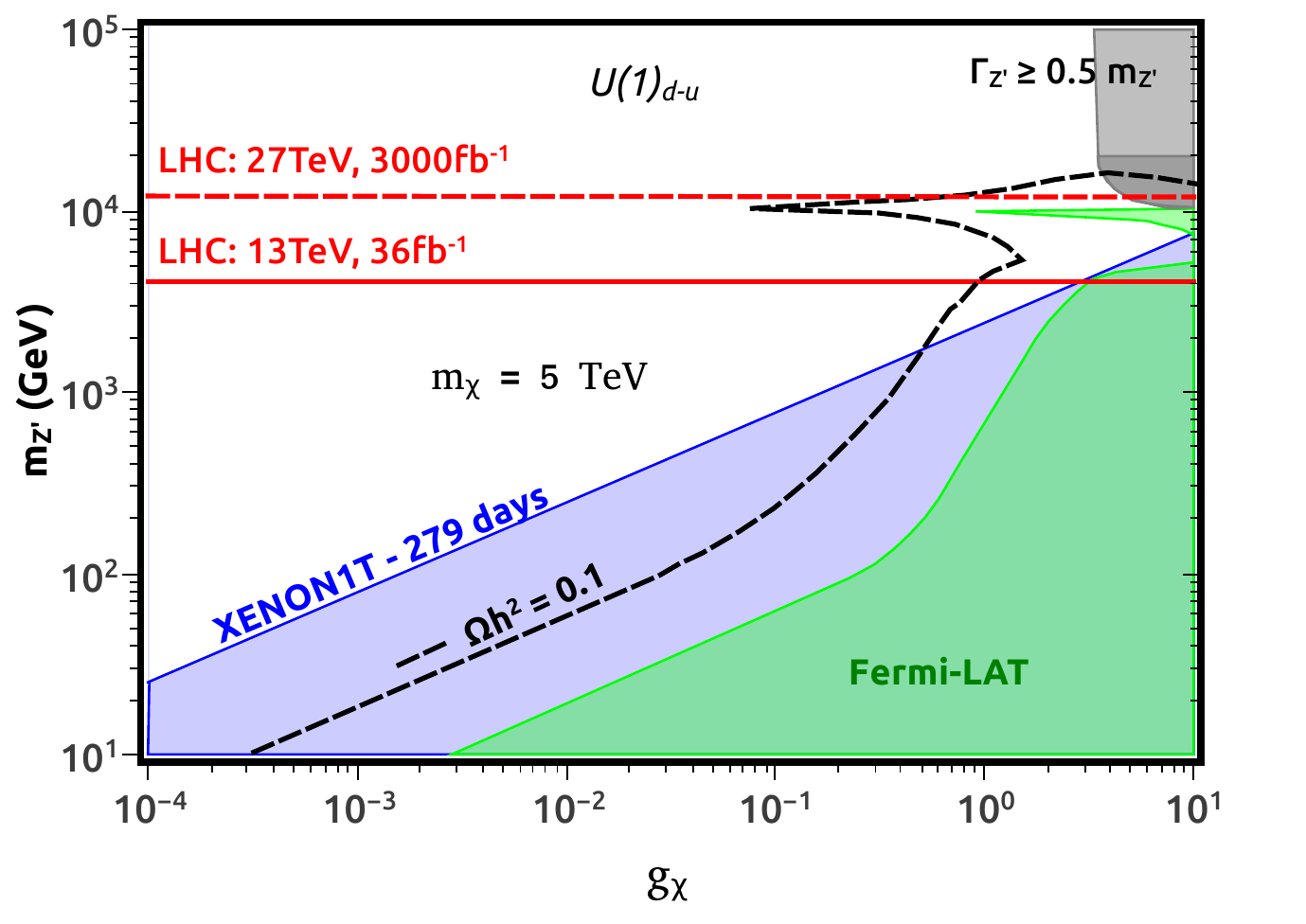}
\caption{
{Same than figure \ref{fig1} but for $m_{\chi}=5$~TeV.}
%Dark matter complementarity plot for $m_{\chi}=5$~TeV. In blue we show the excluded region based on the recent XENON1T bound on the spin-independent scattering cross section. In green, the region ruled out by Fermi-LAT through the observation of dwarf spheroidal galaxies. The red horizontal curve is the LHC limit on the $Z^\prime$ mass using dilepton data at $13$~TeV. The black dashed curve is the delimits the region of parameter space that yields the correct relic entirely via the freeze-out mechanism.
}
\label{fig3}
\end{figure}

\subsection{Indirect Detection}

The relevant observable as far as indirect dark matter detection is the differential gamma-ray flux which depends on the dark matter annihilation cross section, the dark matter mass, and the final states produced in the annihilation. There are two possible annihilation modes. One where the Dirac fermion annihilates into SM fermion pairs and another where $\zp$ pairs are produced, the latter being relevant only when $m_{\chi} \geq m_{\zp}$. The dark matter particles are slow-moving, thus this annihilation process is non-relativistic. That said, annihilation cross sections are

\begin{widetext}
\begin{eqnarray}
\label{annihilation:ff}
\sigma v \left( \chi \bar{\chi} \to f \bar{f} \right) &\approx \frac{n_c \sqrt{1-\frac{m_f^2}{m_\chi^2}}}{2 \pi m_{\zp}^4 \left( 4 m_\chi^2 - m_{\zp}^2 \right)^2} \bigg\{ g_{fa}^2 \Big[ 2 g_{\chi \text{v}}^2 m_{\zp}^4 \left(m_\chi^2 - m_f^2 \right) + g_{\chi a}^2 m_f^2 \left( 4 m_\chi^2 - m_{\zp}^2\right)^2 \Big]
\nonumber \\
& + g_{\chi \text{v}}^2 g_{f \text{v}}^2 m_{\zp}^4 \left( 2 m_\chi^2 + m_f^2 \right) \bigg\}
~,
\end{eqnarray}and 

\begin{eqnarray}
\label{annihilation:ZpZp}
\sigma v \left( \chi \bar{\chi} \to \zp \zp \right) &\approx \frac{1}{16 \pi m_\chi^2 m_{\zp}^2} \left(1 - \frac{m_{\zp}^2}{m_\chi^2}\right)^{3/2} \left(1 - \frac{m_{\zp}^2}{2 m_\chi^2}\right)^{-2}
\nonumber \\
&\times \Big[ 8 g_{\chi \text{v}}^2 g_{\chi a}^2 m_\chi^2 + \left( g_{\chi \text{v}}^4 + g_{\chi a}^4 - 6 g_{\chi \text{v}}^2 g_{\chi a}^2\right) m_{\zp}^2 \Big]
~.
\end{eqnarray}

\end{widetext}where $v$ is the relative velocity of the incoming dark matter particles, $n_c$ is the color number of the final states SM fermions. The Eq.\ref{annihilation:ff} is not valid near the $Z^\prime$ resonance, i.e. when $m_{\zp} \sim 2 m_{\chi}$. In this case, the $\zp$ width should be properly included. We handled this procedure numerically within the MicrOmegas package \cite{Belanger:2006is,Belanger:2008sj,Belanger:2013oya}. With the results we found the region of parameter space that yields that correct dark matter relic density of $\Omega h^2=0.12$ according to Planck \cite{Ade:2015xua}. The region of parameter that furnishes $\Omega h^2=0.12$ is displayed in the figures \ref{fig1}-\ref{fig3} with a dashed black curve.

Concerning indirect detection, there is another quantity to be determined which is the gamma-ray yield. To compute it we implemented the model in Pythia 8 \cite{Sjostrand:2007gs} and took into account all relevant annihilation channels. With the annihilation cross section and energy spectrum at hand, we can compute the dark matter differential flux and compare with upper limits provided by Fermi-LAT \cite{Ackermann:2015zua}. Using this procedure we drew Fermi-LAT exclusion regions displayed in green in figures \ref{fig1}-\ref{fig3}. There are other competitive limits on the dark matter annihilation cross section but we adopted the ones stemming from dwarf spheroidal galaxies for being more robust \cite{Hooper:2012sr,Kong:2014haa,Dutta:2015ysa}. Since the annihilation cross section is s-wave, the indirect detection limits from Fermi-LAT trace the relic density curve as seen in the figures. We highlight that there are bounds resulted from the production of gamma-ray lines at one-loop level but these bounds are not competitive for vector mediators because of the loop suppression and the relative weak experimental sensitivity to gamma-ray lines\cite{Duerr:2015wfa,Profumo:2016idl}.

\section{Collider}

The collider bounds we derived are based on the dilepton data provided ATLAS collaboration using 13TeV center-of-mass energy with about $36~ \mathrm{fb^{-1}}$ of data \cite{Aaboud:2017buh}. Dilepton resonance searches are effective laboratories to constrain the properties of $\zp$ gauge bosons that possess sizable coupling to charged leptons. The couplings to quarks are also sizable but dilepton searches provide stronger bounds simply because the signal suffers from a smaller SM background. We implemented the model in Madgraph and followed the receipt explained in detail in \cite{Camargo:2018klg} to find the lower mass bound $m_{\zp} > 4068$~GeV. This bound is applicable when the invisible $\zp$ decay into dark matter is not accessible. When $m_{\zp} > 2 m_{\chi}$ this invisible decay opens up diminishing the $\zp$ branching ratio into charged lepton and thus weakening the LHC sensitivity. This feature is visible in the figures \ref{fig1} and \ref{fig2} where the limit on $Z'$ mass from LHC searches becomes weaker once $g_\chi = g_{\chi v} = g_{\chi a}$ reaches a reasonable value ($\gtrsim 0.1 - 1$). On the other hand, this is clearly not visible in fig.\ref{fig3} where the dark matter mass (5 TeV) always lies above the dilepton limit on $Z'$ mass.\\

\section{Discussion and Conclusions}

In figures \ref{fig1}-\ref{fig3} we collect all these bounds to find the allowed region of parameter space. It is quite clear how important the recent XENON1T bound is to probe this model. XENON1T excludes alone most of the parameter of the model that yields the correct relic density even for dark matter masses up to $5$~TeV. The indirect detection limits are subdominant in the case of vector mediator simply because of the stringent bounds stemming from LHC on the gauge boson mass. The model only survives for dark matter masses about $\sim 5$~TeV. This happens because the $\zp$ resonance that drives the relic density for $m_{\zp} \sim 2 m_{\chi}$ takes place outside the LHC sensitivity. In summary, the combination of current LHC and XENON1T data pushes the dark $E_6$ model to live at the multi TeV window.\\

Moreover, we also derived the high-energy LHC (HE-LHC) sensitivity using the code provided in \cite{Papucci:2014rja} with high-luminosity. The HE-LHC reach is based on a $27$~TeV proton-proton collision with an integrated luminosity of $3000 fb^{-1}$. In case of null results, this configuration can probe $Z^\prime$ masses up to $12$~TeV. This forecast is exhibited with a dashed red line in figures \ref{fig1}-\ref{fig3}. \\

Considering the prospects for the HE-LHC, we conclude from figures \ref{fig1}-\ref{fig3} that the LHC upgrade is paramount to nearly close the window for a Dirac fermion dark matter in the $E_6$-inspired model described here for dark matter masses below $5$~TeV. Notice that there is tiny region in figure \ref{fig3} that still survives after the HE-LHC prospect is included. We emphasize that this conclusion relies on the fact that the simplified Lagrangian in Eq.\ref{equation:main} provides a good description of the dark matter phenomenology of the dark $E_6$ model and that $g_{\zp}=1$. A smaller $g_{\zp}$ coupling would certainly weaken the collider and the direct detection bounds, however a larger $g_\chi$ would be needed to still reproduce the correct dark matter relic density. At the end, the overall conclusions would remain the same.

\section*{Acknowledgements}

We thank the
High Performance Computing Center (NPAD) at UFRN for providing computational resources. FSQ and DC acknowledge support from MEC and UFRN. FSQ also thanks the financial support from ICTP-SAIFR FAPESP grant 2016/01343-7. Y.M. acknowledges partial support from the European
Union Horizon 2020 research and innovation program under the Marie Sklodowska-Curie:
RISE InvisiblesPlus (grant agreement No 690575), the ITN Elusives (grant agreement No
674896), and the France-US PICS MicroDark.

\bibliography{darkmatter}

\begin{thebibliography}{103}%
\makeatletter
\providecommand \@ifxundefined [1]{%
 \@ifx{#1\undefined}
}%
\providecommand \@ifnum [1]{%
 \ifnum #1\expandafter \@firstoftwo
 \else \expandafter \@secondoftwo
 \fi
}%
\providecommand \@ifx [1]{%
 \ifx #1\expandafter \@firstoftwo
 \else \expandafter \@secondoftwo
 \fi
}%
\providecommand \natexlab [1]{#1}%
\providecommand \enquote  [1]{``#1''}%
\providecommand \bibnamefont  [1]{#1}%
\providecommand \bibfnamefont [1]{#1}%
\providecommand \citenamefont [1]{#1}%
\providecommand \href@noop [0]{\@secondoftwo}%
\providecommand \href [0]{\begingroup \@sanitize@url \@href}%
\providecommand \@href[1]{\@@startlink{#1}\@@href}%
\providecommand \@@href[1]{\endgroup#1\@@endlink}%
\providecommand \@sanitize@url [0]{\catcode `\\12\catcode `\$12\catcode
  `\&12\catcode `\#12\catcode `\^12\catcode `\_12\catcode `\%12\relax}%
\providecommand \@@startlink[1]{}%
\providecommand \@@endlink[0]{}%
\providecommand \url  [0]{\begingroup\@sanitize@url \@url }%
\providecommand \@url [1]{\endgroup\@href {#1}{\urlprefix }}%
\providecommand \urlprefix  [0]{URL }%
\providecommand \Eprint [0]{\href }%
\providecommand \doibase [0]{http://dx.doi.org/}%
\providecommand \selectlanguage [0]{\@gobble}%
\providecommand \bibinfo  [0]{\@secondoftwo}%
\providecommand \bibfield  [0]{\@secondoftwo}%
\providecommand \translation [1]{[#1]}%
\providecommand \BibitemOpen [0]{}%
\providecommand \bibitemStop [0]{}%
\providecommand \bibitemNoStop [0]{.\EOS\space}%
\providecommand \EOS [0]{\spacefactor3000\relax}%
\providecommand \BibitemShut  [1]{\csname bibitem#1\endcsname}%
\let\auto@bib@innerbib\@empty
%</preamble>
\bibitem [{\citenamefont {Bertone}\ \emph {et~al.}(2005)\citenamefont
  {Bertone}, \citenamefont {Hooper},\ and\ \citenamefont
  {Silk}}]{Bertone:2004pz}%
  \BibitemOpen
  \bibfield  {author} {\bibinfo {author} {\bibfnamefont {G.}~\bibnamefont
  {Bertone}}, \bibinfo {author} {\bibfnamefont {D.}~\bibnamefont {Hooper}}, \
  and\ \bibinfo {author} {\bibfnamefont {J.}~\bibnamefont {Silk}},\ }\href
  {\doibase 10.1016/j.physrep.2004.08.031} {\bibfield  {journal} {\bibinfo
  {journal} {Phys. Rept.}\ }\textbf {\bibinfo {volume} {405}},\ \bibinfo
  {pages} {279} (\bibinfo {year} {2005})},\ \Eprint
  {http://arxiv.org/abs/hep-ph/0404175} {arXiv:hep-ph/0404175 [hep-ph]}
  \BibitemShut {NoStop}%
%%CITATION = HEP-PH/0404175;%%
\bibitem [{\citenamefont {Queiroz}\ \emph {et~al.}(2017)\citenamefont
  {Queiroz}, \citenamefont {Rodejohann},\ and\ \citenamefont
  {Yaguna}}]{Queiroz:2016sxf}%
  \BibitemOpen
  \bibfield  {author} {\bibinfo {author} {\bibfnamefont {F.~S.}\ \bibnamefont
  {Queiroz}}, \bibinfo {author} {\bibfnamefont {W.}~\bibnamefont {Rodejohann}},
  \ and\ \bibinfo {author} {\bibfnamefont {C.~E.}\ \bibnamefont {Yaguna}},\
  }\href {\doibase 10.1103/PhysRevD.95.095010} {\bibfield  {journal} {\bibinfo
  {journal} {Phys. Rev.}\ }\textbf {\bibinfo {volume} {D95}},\ \bibinfo {pages}
  {095010} (\bibinfo {year} {2017})},\ \Eprint
  {http://arxiv.org/abs/1610.06581} {arXiv:1610.06581 [hep-ph]} \BibitemShut
  {NoStop}%
%%CITATION = ARXIV:1610.06581;%%
\bibitem [{\citenamefont {Kavanagh}\ \emph {et~al.}(2017)\citenamefont
  {Kavanagh}, \citenamefont {Queiroz}, \citenamefont {Rodejohann},\ and\
  \citenamefont {Yaguna}}]{Kavanagh:2017hcl}%
  \BibitemOpen
  \bibfield  {author} {\bibinfo {author} {\bibfnamefont {B.~J.}\ \bibnamefont
  {Kavanagh}}, \bibinfo {author} {\bibfnamefont {F.~S.}\ \bibnamefont
  {Queiroz}}, \bibinfo {author} {\bibfnamefont {W.}~\bibnamefont {Rodejohann}},
  \ and\ \bibinfo {author} {\bibfnamefont {C.~E.}\ \bibnamefont {Yaguna}},\
  }\href@noop {} {\  (\bibinfo {year} {2017})},\ \Eprint
  {http://arxiv.org/abs/1706.07819} {arXiv:1706.07819 [hep-ph]} \BibitemShut
  {NoStop}%
%%CITATION = ARXIV:1706.07819;%%
\bibitem [{\citenamefont {Bergstrom}(2009)}]{Bergstrom:2009ib}%
  \BibitemOpen
  \bibfield  {author} {\bibinfo {author} {\bibfnamefont {L.}~\bibnamefont
  {Bergstrom}},\ }\href {\doibase 10.1088/1367-2630/11/10/105006} {\bibfield
  {journal} {\bibinfo  {journal} {New J. Phys.}\ }\textbf {\bibinfo {volume}
  {11}},\ \bibinfo {pages} {105006} (\bibinfo {year} {2009})},\ \Eprint
  {http://arxiv.org/abs/0903.4849} {arXiv:0903.4849 [hep-ph]} \BibitemShut
  {NoStop}%
%%CITATION = ARXIV:0903.4849;%%
\bibitem [{\citenamefont {Klasen}\ \emph {et~al.}(2015)\citenamefont {Klasen},
  \citenamefont {Pohl},\ and\ \citenamefont {Sigl}}]{Klasen:2015uma}%
  \BibitemOpen
  \bibfield  {author} {\bibinfo {author} {\bibfnamefont {M.}~\bibnamefont
  {Klasen}}, \bibinfo {author} {\bibfnamefont {M.}~\bibnamefont {Pohl}}, \ and\
  \bibinfo {author} {\bibfnamefont {G.}~\bibnamefont {Sigl}},\ }\href {\doibase
  10.1016/j.ppnp.2015.07.001} {\bibfield  {journal} {\bibinfo  {journal} {Prog.
  Part. Nucl. Phys.}\ }\textbf {\bibinfo {volume} {85}},\ \bibinfo {pages} {1}
  (\bibinfo {year} {2015})},\ \Eprint {http://arxiv.org/abs/1507.03800}
  {arXiv:1507.03800 [hep-ph]} \BibitemShut {NoStop}%
%%CITATION = ARXIV:1507.03800;%%
\bibitem [{\citenamefont {Jungman}\ \emph {et~al.}(1996)\citenamefont
  {Jungman}, \citenamefont {Kamionkowski},\ and\ \citenamefont
  {Griest}}]{Jungman:1995df}%
  \BibitemOpen
  \bibfield  {author} {\bibinfo {author} {\bibfnamefont {G.}~\bibnamefont
  {Jungman}}, \bibinfo {author} {\bibfnamefont {M.}~\bibnamefont
  {Kamionkowski}}, \ and\ \bibinfo {author} {\bibfnamefont {K.}~\bibnamefont
  {Griest}},\ }\href {\doibase 10.1016/0370-1573(95)00058-5} {\bibfield
  {journal} {\bibinfo  {journal} {Phys.Rept.}\ }\textbf {\bibinfo {volume}
  {267}},\ \bibinfo {pages} {195} (\bibinfo {year} {1996})},\ \Eprint
  {http://arxiv.org/abs/hep-ph/9506380} {arXiv:hep-ph/9506380 [hep-ph]}
  \BibitemShut {NoStop}%
%%CITATION = HEP-PH/9506380;%%
\bibitem [{\citenamefont {Baer}\ \emph {et~al.}(2015)\citenamefont {Baer},
  \citenamefont {Choi}, \citenamefont {Kim},\ and\ \citenamefont
  {Roszkowski}}]{Baer:2014eja}%
  \BibitemOpen
  \bibfield  {author} {\bibinfo {author} {\bibfnamefont {H.}~\bibnamefont
  {Baer}}, \bibinfo {author} {\bibfnamefont {K.-Y.}\ \bibnamefont {Choi}},
  \bibinfo {author} {\bibfnamefont {J.~E.}\ \bibnamefont {Kim}}, \ and\
  \bibinfo {author} {\bibfnamefont {L.}~\bibnamefont {Roszkowski}},\ }\href
  {\doibase 10.1016/j.physrep.2014.10.002} {\bibfield  {journal} {\bibinfo
  {journal} {Phys. Rept.}\ }\textbf {\bibinfo {volume} {555}},\ \bibinfo
  {pages} {1} (\bibinfo {year} {2015})},\ \Eprint
  {http://arxiv.org/abs/1407.0017} {arXiv:1407.0017 [hep-ph]} \BibitemShut
  {NoStop}%
%%CITATION = ARXIV:1407.0017;%%
\bibitem [{\citenamefont {Arcadi}\ \emph
  {et~al.}(2017{\natexlab{a}})\citenamefont {Arcadi}, \citenamefont {Dutra},
  \citenamefont {Ghosh}, \citenamefont {Lindner}, \citenamefont {Mambrini},
  \citenamefont {Pierre}, \citenamefont {Profumo},\ and\ \citenamefont
  {Queiroz}}]{Arcadi:2017kky}%
  \BibitemOpen
  \bibfield  {author} {\bibinfo {author} {\bibfnamefont {G.}~\bibnamefont
  {Arcadi}}, \bibinfo {author} {\bibfnamefont {M.}~\bibnamefont {Dutra}},
  \bibinfo {author} {\bibfnamefont {P.}~\bibnamefont {Ghosh}}, \bibinfo
  {author} {\bibfnamefont {M.}~\bibnamefont {Lindner}}, \bibinfo {author}
  {\bibfnamefont {Y.}~\bibnamefont {Mambrini}}, \bibinfo {author}
  {\bibfnamefont {M.}~\bibnamefont {Pierre}}, \bibinfo {author} {\bibfnamefont
  {S.}~\bibnamefont {Profumo}}, \ and\ \bibinfo {author} {\bibfnamefont
  {F.~S.}\ \bibnamefont {Queiroz}},\ }\href@noop {} {\  (\bibinfo {year}
  {2017}{\natexlab{a}})},\ \Eprint {http://arxiv.org/abs/1703.07364}
  {arXiv:1703.07364 [hep-ph]} \BibitemShut {NoStop}%
%%CITATION = ARXIV:1703.07364;%%
\bibitem [{\citenamefont {Cushman}\ \emph {et~al.}(2013)\citenamefont {Cushman}
  \emph {et~al.}}]{Cushman:2013zza}%
  \BibitemOpen
  \bibfield  {author} {\bibinfo {author} {\bibfnamefont {P.}~\bibnamefont
  {Cushman}} \emph {et~al.},\ }in\ \href
  {http://inspirehep.net/record/1262767/files/arXiv:1310.8327.pdf} {\emph
  {\bibinfo {booktitle} {{Community Summer Study 2013: Snowmass on the
  Mississippi (CSS2013) Minneapolis, MN, USA, July 29-August 6, 2013}}}}\
  (\bibinfo {year} {2013})\ \Eprint {http://arxiv.org/abs/1310.8327}
  {arXiv:1310.8327 [hep-ex]} \BibitemShut {NoStop}%
%%CITATION = ARXIV:1310.8327;%%
\bibitem [{\citenamefont {Queiroz}(2016)}]{Queiroz:2016awc}%
  \BibitemOpen
  \bibfield  {author} {\bibinfo {author} {\bibfnamefont {F.~S.}\ \bibnamefont
  {Queiroz}},\ }in\ \href
  {http://inspirehep.net/record/1466249/files/arXiv:1605.08788.pdf} {\emph
  {\bibinfo {booktitle} {{Proceedings, 51st Rencontres de Moriond on
  Electroweak Interactions and Unified Theories: La Thuile, Italy, March 12-19,
  2016}}}},\ \bibinfo {organization} {ARISF}\ (\bibinfo  {publisher} {ARISF},\
  \bibinfo {year} {2016})\ pp.\ \bibinfo {pages} {427--436},\ \Eprint
  {http://arxiv.org/abs/1605.08788} {arXiv:1605.08788 [hep-ph]} \BibitemShut
  {NoStop}%
%%CITATION = ARXIV:1605.08788;%%
\bibitem [{\citenamefont {Marrodán~Undagoitia}\ and\ \citenamefont
  {Rauch}(2016)}]{Undagoitia:2015gya}%
  \BibitemOpen
  \bibfield  {author} {\bibinfo {author} {\bibfnamefont {T.}~\bibnamefont
  {Marrodán~Undagoitia}}\ and\ \bibinfo {author} {\bibfnamefont
  {L.}~\bibnamefont {Rauch}},\ }\href {\doibase 10.1088/0954-3899/43/1/013001}
  {\bibfield  {journal} {\bibinfo  {journal} {J. Phys.}\ }\textbf {\bibinfo
  {volume} {G43}},\ \bibinfo {pages} {013001} (\bibinfo {year} {2016})},\
  \Eprint {http://arxiv.org/abs/1509.08767} {arXiv:1509.08767
  [physics.ins-det]} \BibitemShut {NoStop}%
%%CITATION = ARXIV:1509.08767;%%
\bibitem [{\citenamefont {Gondolo}(1996)}]{Gondolo:1996qw}%
  \BibitemOpen
  \bibfield  {author} {\bibinfo {author} {\bibfnamefont {P.}~\bibnamefont
  {Gondolo}},\ }in\ \href
  {http://inspirehep.net/record/418555/files/Pages_from_C96-01-20_41.pdf}
  {\emph {\bibinfo {booktitle} {{Dark matter in cosmology, quantum
  measurements, experimental gravitation. Proceedings, 31st Rencontres de
  Moriond, 16th Moriond Workshop, Les Arcs, France, January 2-27, 1996}}}}\
  (\bibinfo {year} {1996})\ pp.\ \bibinfo {pages} {41--51},\ \Eprint
  {http://arxiv.org/abs/hep-ph/9605290} {arXiv:hep-ph/9605290 [hep-ph]}
  \BibitemShut {NoStop}%
%%CITATION = HEP-PH/9605290;%%
\bibitem [{\citenamefont {Pato}\ \emph {et~al.}(2011)\citenamefont {Pato},
  \citenamefont {Baudis}, \citenamefont {Bertone}, \citenamefont {Ruiz~de
  Austri}, \citenamefont {Strigari},\ and\ \citenamefont
  {Trotta}}]{Pato:2010zk}%
  \BibitemOpen
  \bibfield  {author} {\bibinfo {author} {\bibfnamefont {M.}~\bibnamefont
  {Pato}}, \bibinfo {author} {\bibfnamefont {L.}~\bibnamefont {Baudis}},
  \bibinfo {author} {\bibfnamefont {G.}~\bibnamefont {Bertone}}, \bibinfo
  {author} {\bibfnamefont {R.}~\bibnamefont {Ruiz~de Austri}}, \bibinfo
  {author} {\bibfnamefont {L.~E.}\ \bibnamefont {Strigari}}, \ and\ \bibinfo
  {author} {\bibfnamefont {R.}~\bibnamefont {Trotta}},\ }\href {\doibase
  10.1103/PhysRevD.83.083505} {\bibfield  {journal} {\bibinfo  {journal} {Phys.
  Rev.}\ }\textbf {\bibinfo {volume} {D83}},\ \bibinfo {pages} {083505}
  (\bibinfo {year} {2011})},\ \Eprint {http://arxiv.org/abs/1012.3458}
  {arXiv:1012.3458 [astro-ph.CO]} \BibitemShut {NoStop}%
%%CITATION = ARXIV:1012.3458;%%
\bibitem [{\citenamefont {Freese}\ \emph {et~al.}(2013)\citenamefont {Freese},
  \citenamefont {Lisanti},\ and\ \citenamefont {Savage}}]{Freese:2012xd}%
  \BibitemOpen
  \bibfield  {author} {\bibinfo {author} {\bibfnamefont {K.}~\bibnamefont
  {Freese}}, \bibinfo {author} {\bibfnamefont {M.}~\bibnamefont {Lisanti}}, \
  and\ \bibinfo {author} {\bibfnamefont {C.}~\bibnamefont {Savage}},\ }\href
  {\doibase 10.1103/RevModPhys.85.1561} {\bibfield  {journal} {\bibinfo
  {journal} {Rev. Mod. Phys.}\ }\textbf {\bibinfo {volume} {85}},\ \bibinfo
  {pages} {1561} (\bibinfo {year} {2013})},\ \Eprint
  {http://arxiv.org/abs/1209.3339} {arXiv:1209.3339 [astro-ph.CO]} \BibitemShut
  {NoStop}%
%%CITATION = ARXIV:1209.3339;%%
\bibitem [{\citenamefont {Del~Nobile}\ \emph {et~al.}(2015)\citenamefont
  {Del~Nobile}, \citenamefont {Gelmini}, \citenamefont {Gondolo},\ and\
  \citenamefont {Huh}}]{DelNobile:2014sja}%
  \BibitemOpen
  \bibfield  {author} {\bibinfo {author} {\bibfnamefont {E.}~\bibnamefont
  {Del~Nobile}}, \bibinfo {author} {\bibfnamefont {G.~B.}\ \bibnamefont
  {Gelmini}}, \bibinfo {author} {\bibfnamefont {P.}~\bibnamefont {Gondolo}}, \
  and\ \bibinfo {author} {\bibfnamefont {J.-H.}\ \bibnamefont {Huh}},\
  }\bibfield  {booktitle} {\emph {\bibinfo {booktitle} {{Proceedings, 13th
  International Conference on Topics in Astroparticle and Underground Physics
  (TAUP 2013): Asilomar, California, September 8-13, 2013}}},\ }\href {\doibase
  10.1016/j.phpro.2014.12.009} {\bibfield  {journal} {\bibinfo  {journal}
  {Phys. Procedia}\ }\textbf {\bibinfo {volume} {61}},\ \bibinfo {pages} {45}
  (\bibinfo {year} {2015})},\ \Eprint {http://arxiv.org/abs/1405.5582}
  {arXiv:1405.5582 [hep-ph]} \BibitemShut {NoStop}%
%%CITATION = ARXIV:1405.5582;%%
\bibitem [{\citenamefont {Catena}\ and\ \citenamefont
  {Ullio}(2010)}]{Catena:2009mf}%
  \BibitemOpen
  \bibfield  {author} {\bibinfo {author} {\bibfnamefont {R.}~\bibnamefont
  {Catena}}\ and\ \bibinfo {author} {\bibfnamefont {P.}~\bibnamefont {Ullio}},\
  }\href {\doibase 10.1088/1475-7516/2010/08/004} {\bibfield  {journal}
  {\bibinfo  {journal} {JCAP}\ }\textbf {\bibinfo {volume} {1008}},\ \bibinfo
  {pages} {004} (\bibinfo {year} {2010})},\ \Eprint
  {http://arxiv.org/abs/0907.0018} {arXiv:0907.0018 [astro-ph.CO]} \BibitemShut
  {NoStop}%
%%CITATION = ARXIV:0907.0018;%%
\bibitem [{\citenamefont {Pato}\ \emph {et~al.}(2010)\citenamefont {Pato},
  \citenamefont {Agertz}, \citenamefont {Bertone}, \citenamefont {Moore},\ and\
  \citenamefont {Teyssier}}]{Pato:2010yq}%
  \BibitemOpen
  \bibfield  {author} {\bibinfo {author} {\bibfnamefont {M.}~\bibnamefont
  {Pato}}, \bibinfo {author} {\bibfnamefont {O.}~\bibnamefont {Agertz}},
  \bibinfo {author} {\bibfnamefont {G.}~\bibnamefont {Bertone}}, \bibinfo
  {author} {\bibfnamefont {B.}~\bibnamefont {Moore}}, \ and\ \bibinfo {author}
  {\bibfnamefont {R.}~\bibnamefont {Teyssier}},\ }\href {\doibase
  10.1103/PhysRevD.82.023531} {\bibfield  {journal} {\bibinfo  {journal} {Phys.
  Rev.}\ }\textbf {\bibinfo {volume} {D82}},\ \bibinfo {pages} {023531}
  (\bibinfo {year} {2010})},\ \Eprint {http://arxiv.org/abs/1006.1322}
  {arXiv:1006.1322 [astro-ph.HE]} \BibitemShut {NoStop}%
%%CITATION = ARXIV:1006.1322;%%
\bibitem [{\citenamefont {Iocco}\ \emph {et~al.}(2011)\citenamefont {Iocco},
  \citenamefont {Pato}, \citenamefont {Bertone},\ and\ \citenamefont
  {Jetzer}}]{Iocco:2011jz}%
  \BibitemOpen
  \bibfield  {author} {\bibinfo {author} {\bibfnamefont {F.}~\bibnamefont
  {Iocco}}, \bibinfo {author} {\bibfnamefont {M.}~\bibnamefont {Pato}},
  \bibinfo {author} {\bibfnamefont {G.}~\bibnamefont {Bertone}}, \ and\
  \bibinfo {author} {\bibfnamefont {P.}~\bibnamefont {Jetzer}},\ }\href
  {\doibase 10.1088/1475-7516/2011/11/029} {\bibfield  {journal} {\bibinfo
  {journal} {JCAP}\ }\textbf {\bibinfo {volume} {1111}},\ \bibinfo {pages}
  {029} (\bibinfo {year} {2011})},\ \Eprint {http://arxiv.org/abs/1107.5810}
  {arXiv:1107.5810 [astro-ph.GA]} \BibitemShut {NoStop}%
%%CITATION = ARXIV:1107.5810;%%
\bibitem [{\citenamefont {Kelso}\ \emph {et~al.}(2012)\citenamefont {Kelso},
  \citenamefont {Hooper},\ and\ \citenamefont {Buckley}}]{Kelso:2011gd}%
  \BibitemOpen
  \bibfield  {author} {\bibinfo {author} {\bibfnamefont {C.}~\bibnamefont
  {Kelso}}, \bibinfo {author} {\bibfnamefont {D.}~\bibnamefont {Hooper}}, \
  and\ \bibinfo {author} {\bibfnamefont {M.~R.}\ \bibnamefont {Buckley}},\
  }\href {\doibase 10.1103/PhysRevD.85.043515} {\bibfield  {journal} {\bibinfo
  {journal} {Phys. Rev.}\ }\textbf {\bibinfo {volume} {D85}},\ \bibinfo {pages}
  {043515} (\bibinfo {year} {2012})},\ \Eprint {http://arxiv.org/abs/1110.5338}
  {arXiv:1110.5338 [astro-ph.CO]} \BibitemShut {NoStop}%
%%CITATION = ARXIV:1110.5338;%%
\bibitem [{\citenamefont {Akerib}\ \emph {et~al.}(2014)\citenamefont {Akerib}
  \emph {et~al.}}]{Akerib:2013tjd}%
  \BibitemOpen
  \bibfield  {author} {\bibinfo {author} {\bibfnamefont {D.~S.}\ \bibnamefont
  {Akerib}} \emph {et~al.} (\bibinfo {collaboration} {LUX}),\ }\href {\doibase
  10.1103/PhysRevLett.112.091303} {\bibfield  {journal} {\bibinfo  {journal}
  {Phys. Rev. Lett.}\ }\textbf {\bibinfo {volume} {112}},\ \bibinfo {pages}
  {091303} (\bibinfo {year} {2014})},\ \Eprint {http://arxiv.org/abs/1310.8214}
  {arXiv:1310.8214 [astro-ph.CO]} \BibitemShut {NoStop}%
%%CITATION = ARXIV:1310.8214;%%
\bibitem [{\citenamefont {Agnese}\ \emph {et~al.}(2014)\citenamefont {Agnese}
  \emph {et~al.}}]{Agnese:2013jaa}%
  \BibitemOpen
  \bibfield  {author} {\bibinfo {author} {\bibfnamefont {R.}~\bibnamefont
  {Agnese}} \emph {et~al.} (\bibinfo {collaboration} {SuperCDMS}),\ }\href
  {\doibase 10.1103/PhysRevLett.112.041302} {\bibfield  {journal} {\bibinfo
  {journal} {Phys. Rev. Lett.}\ }\textbf {\bibinfo {volume} {112}},\ \bibinfo
  {pages} {041302} (\bibinfo {year} {2014})},\ \Eprint
  {http://arxiv.org/abs/1309.3259} {arXiv:1309.3259 [physics.ins-det]}
  \BibitemShut {NoStop}%
%%CITATION = ARXIV:1309.3259;%%
\bibitem [{\citenamefont {Scott}\ \emph {et~al.}(2010)\citenamefont {Scott},
  \citenamefont {Conrad}, \citenamefont {Edsjo}, \citenamefont {Bergstrom},
  \citenamefont {Farnier},\ and\ \citenamefont {Akrami}}]{Scott:2009jn}%
  \BibitemOpen
  \bibfield  {author} {\bibinfo {author} {\bibfnamefont {P.}~\bibnamefont
  {Scott}}, \bibinfo {author} {\bibfnamefont {J.}~\bibnamefont {Conrad}},
  \bibinfo {author} {\bibfnamefont {J.}~\bibnamefont {Edsjo}}, \bibinfo
  {author} {\bibfnamefont {L.}~\bibnamefont {Bergstrom}}, \bibinfo {author}
  {\bibfnamefont {C.}~\bibnamefont {Farnier}}, \ and\ \bibinfo {author}
  {\bibfnamefont {Y.}~\bibnamefont {Akrami}},\ }\href {\doibase
  10.1088/1475-7516/2010/01/031} {\bibfield  {journal} {\bibinfo  {journal}
  {JCAP}\ }\textbf {\bibinfo {volume} {1001}},\ \bibinfo {pages} {031}
  (\bibinfo {year} {2010})},\ \Eprint {http://arxiv.org/abs/0909.3300}
  {arXiv:0909.3300 [astro-ph.CO]} \BibitemShut {NoStop}%
%%CITATION = ARXIV:0909.3300;%%
\bibitem [{\citenamefont {Abdo}\ \emph {et~al.}(2010)\citenamefont {Abdo} \emph
  {et~al.}}]{Abdo:2010ex}%
  \BibitemOpen
  \bibfield  {author} {\bibinfo {author} {\bibfnamefont {A.~A.}\ \bibnamefont
  {Abdo}} \emph {et~al.} (\bibinfo {collaboration} {Fermi-LAT}),\ }\href
  {\doibase 10.1088/0004-637X/712/1/147} {\bibfield  {journal} {\bibinfo
  {journal} {Astrophys. J.}\ }\textbf {\bibinfo {volume} {712}},\ \bibinfo
  {pages} {147} (\bibinfo {year} {2010})},\ \Eprint
  {http://arxiv.org/abs/1001.4531} {arXiv:1001.4531 [astro-ph.CO]} \BibitemShut
  {NoStop}%
%%CITATION = ARXIV:1001.4531;%%
\bibitem [{\citenamefont {Hooper}\ \emph {et~al.}(2013)\citenamefont {Hooper},
  \citenamefont {Kelso},\ and\ \citenamefont {Queiroz}}]{Hooper:2012sr}%
  \BibitemOpen
  \bibfield  {author} {\bibinfo {author} {\bibfnamefont {D.}~\bibnamefont
  {Hooper}}, \bibinfo {author} {\bibfnamefont {C.}~\bibnamefont {Kelso}}, \
  and\ \bibinfo {author} {\bibfnamefont {F.~S.}\ \bibnamefont {Queiroz}},\
  }\href {\doibase 10.1016/j.astropartphys.2013.04.007} {\bibfield  {journal}
  {\bibinfo  {journal} {Astropart. Phys.}\ }\textbf {\bibinfo {volume} {46}},\
  \bibinfo {pages} {55} (\bibinfo {year} {2013})},\ \Eprint
  {http://arxiv.org/abs/1209.3015} {arXiv:1209.3015 [astro-ph.HE]} \BibitemShut
  {NoStop}%
%%CITATION = ARXIV:1209.3015;%%
\bibitem [{\citenamefont {Gonzalez-Morales}\ \emph {et~al.}(2014)\citenamefont
  {Gonzalez-Morales}, \citenamefont {Profumo},\ and\ \citenamefont
  {Queiroz}}]{Gonzalez-Morales:2014eaa}%
  \BibitemOpen
  \bibfield  {author} {\bibinfo {author} {\bibfnamefont {A.~X.}\ \bibnamefont
  {Gonzalez-Morales}}, \bibinfo {author} {\bibfnamefont {S.}~\bibnamefont
  {Profumo}}, \ and\ \bibinfo {author} {\bibfnamefont {F.~S.}\ \bibnamefont
  {Queiroz}},\ }\href {\doibase 10.1103/PhysRevD.90.103508} {\bibfield
  {journal} {\bibinfo  {journal} {Phys. Rev.}\ }\textbf {\bibinfo {volume}
  {D90}},\ \bibinfo {pages} {103508} (\bibinfo {year} {2014})},\ \Eprint
  {http://arxiv.org/abs/1406.2424} {arXiv:1406.2424 [astro-ph.HE]} \BibitemShut
  {NoStop}%
%%CITATION = ARXIV:1406.2424;%%
\bibitem [{\citenamefont {Mambrini}\ \emph {et~al.}(2016)\citenamefont
  {Mambrini}, \citenamefont {Profumo},\ and\ \citenamefont
  {Queiroz}}]{Mambrini:2015sia}%
  \BibitemOpen
  \bibfield  {author} {\bibinfo {author} {\bibfnamefont {Y.}~\bibnamefont
  {Mambrini}}, \bibinfo {author} {\bibfnamefont {S.}~\bibnamefont {Profumo}}, \
  and\ \bibinfo {author} {\bibfnamefont {F.~S.}\ \bibnamefont {Queiroz}},\
  }\href {\doibase 10.1016/j.physletb.2016.07.076} {\bibfield  {journal}
  {\bibinfo  {journal} {Phys. Lett.}\ }\textbf {\bibinfo {volume} {B760}},\
  \bibinfo {pages} {807} (\bibinfo {year} {2016})},\ \Eprint
  {http://arxiv.org/abs/1508.06635} {arXiv:1508.06635 [hep-ph]} \BibitemShut
  {NoStop}%
%%CITATION = ARXIV:1508.06635;%%
\bibitem [{\citenamefont {Baring}\ \emph {et~al.}(2015)\citenamefont {Baring},
  \citenamefont {Ghosh}, \citenamefont {Queiroz},\ and\ \citenamefont
  {Sinha}}]{Baring:2015sza}%
  \BibitemOpen
  \bibfield  {author} {\bibinfo {author} {\bibfnamefont {M.~G.}\ \bibnamefont
  {Baring}}, \bibinfo {author} {\bibfnamefont {T.}~\bibnamefont {Ghosh}},
  \bibinfo {author} {\bibfnamefont {F.~S.}\ \bibnamefont {Queiroz}}, \ and\
  \bibinfo {author} {\bibfnamefont {K.}~\bibnamefont {Sinha}},\ }\href@noop {}
  {\  (\bibinfo {year} {2015})},\ \Eprint {http://arxiv.org/abs/1510.00389}
  {arXiv:1510.00389 [hep-ph]} \BibitemShut {NoStop}%
%%CITATION = ARXIV:1510.00389;%%
\bibitem [{\citenamefont {Li}\ \emph {et~al.}(2016)\citenamefont {Li},
  \citenamefont {Liang}, \citenamefont {Duan}, \citenamefont {Shen},
  \citenamefont {Huang}, \citenamefont {Li}, \citenamefont {Fan}, \citenamefont
  {Liao}, \citenamefont {Feng},\ and\ \citenamefont {Chang}}]{Li:2015kag}%
  \BibitemOpen
  \bibfield  {author} {\bibinfo {author} {\bibfnamefont {S.}~\bibnamefont
  {Li}}, \bibinfo {author} {\bibfnamefont {Y.-F.}\ \bibnamefont {Liang}},
  \bibinfo {author} {\bibfnamefont {K.-K.}\ \bibnamefont {Duan}}, \bibinfo
  {author} {\bibfnamefont {Z.-Q.}\ \bibnamefont {Shen}}, \bibinfo {author}
  {\bibfnamefont {X.}~\bibnamefont {Huang}}, \bibinfo {author} {\bibfnamefont
  {X.}~\bibnamefont {Li}}, \bibinfo {author} {\bibfnamefont {Y.-Z.}\
  \bibnamefont {Fan}}, \bibinfo {author} {\bibfnamefont {N.-H.}\ \bibnamefont
  {Liao}}, \bibinfo {author} {\bibfnamefont {L.}~\bibnamefont {Feng}}, \ and\
  \bibinfo {author} {\bibfnamefont {J.}~\bibnamefont {Chang}},\ }\href
  {\doibase 10.1103/PhysRevD.93.043518} {\bibfield  {journal} {\bibinfo
  {journal} {Phys. Rev.}\ }\textbf {\bibinfo {volume} {D93}},\ \bibinfo {pages}
  {043518} (\bibinfo {year} {2016})},\ \Eprint
  {http://arxiv.org/abs/1511.09252} {arXiv:1511.09252 [astro-ph.HE]}
  \BibitemShut {NoStop}%
%%CITATION = ARXIV:1511.09252;%%
\bibitem [{\citenamefont {Drlica-Wagner}\ \emph {et~al.}(2015)\citenamefont
  {Drlica-Wagner} \emph {et~al.}}]{Drlica-Wagner:2015xua}%
  \BibitemOpen
  \bibfield  {author} {\bibinfo {author} {\bibfnamefont {A.}~\bibnamefont
  {Drlica-Wagner}} \emph {et~al.} (\bibinfo {collaboration} {DES, Fermi-LAT}),\
  }\href {\doibase 10.1088/2041-8205/809/1/L4} {\bibfield  {journal} {\bibinfo
  {journal} {Astrophys. J.}\ }\textbf {\bibinfo {volume} {809}},\ \bibinfo
  {pages} {L4} (\bibinfo {year} {2015})},\ \Eprint
  {http://arxiv.org/abs/1503.02632} {arXiv:1503.02632 [astro-ph.HE]}
  \BibitemShut {NoStop}%
%%CITATION = ARXIV:1503.02632;%%
\bibitem [{\citenamefont {Zhao}\ \emph {et~al.}(2016)\citenamefont {Zhao},
  \citenamefont {Bi}, \citenamefont {Jia}, \citenamefont {Yin},\ and\
  \citenamefont {Zhu}}]{Zhao:2016xie}%
  \BibitemOpen
  \bibfield  {author} {\bibinfo {author} {\bibfnamefont {Y.}~\bibnamefont
  {Zhao}}, \bibinfo {author} {\bibfnamefont {X.-J.}\ \bibnamefont {Bi}},
  \bibinfo {author} {\bibfnamefont {H.-Y.}\ \bibnamefont {Jia}}, \bibinfo
  {author} {\bibfnamefont {P.-F.}\ \bibnamefont {Yin}}, \ and\ \bibinfo
  {author} {\bibfnamefont {F.-R.}\ \bibnamefont {Zhu}},\ }\href {\doibase
  10.1103/PhysRevD.93.083513} {\bibfield  {journal} {\bibinfo  {journal} {Phys.
  Rev.}\ }\textbf {\bibinfo {volume} {D93}},\ \bibinfo {pages} {083513}
  (\bibinfo {year} {2016})},\ \Eprint {http://arxiv.org/abs/1601.02181}
  {arXiv:1601.02181 [astro-ph.HE]} \BibitemShut {NoStop}%
%%CITATION = ARXIV:1601.02181;%%
\bibitem [{\citenamefont {Garcia-Cely}\ and\ \citenamefont
  {Heeck}(2016)}]{Garcia-Cely:2016pse}%
  \BibitemOpen
  \bibfield  {author} {\bibinfo {author} {\bibfnamefont {C.}~\bibnamefont
  {Garcia-Cely}}\ and\ \bibinfo {author} {\bibfnamefont {J.}~\bibnamefont
  {Heeck}},\ }\href {\doibase 10.1088/1475-7516/2016/08/023} {\bibfield
  {journal} {\bibinfo  {journal} {JCAP}\ }\textbf {\bibinfo {volume} {1608}},\
  \bibinfo {pages} {023} (\bibinfo {year} {2016})},\ \Eprint
  {http://arxiv.org/abs/1605.08049} {arXiv:1605.08049 [hep-ph]} \BibitemShut
  {NoStop}%
%%CITATION = ARXIV:1605.08049;%%
\bibitem [{\citenamefont {Li}\ \emph {et~al.}(2018)\citenamefont {Li} \emph
  {et~al.}}]{Li:2018kgy}%
  \BibitemOpen
  \bibfield  {author} {\bibinfo {author} {\bibfnamefont {S.}~\bibnamefont {Li}}
  \emph {et~al.},\ }\href@noop {} {\  (\bibinfo {year} {2018})},\ \Eprint
  {http://arxiv.org/abs/1805.06612} {arXiv:1805.06612 [astro-ph.HE]}
  \BibitemShut {NoStop}%
%%CITATION = ARXIV:1805.06612;%%
\bibitem [{\citenamefont {Ackermann}\ \emph {et~al.}(2015)\citenamefont
  {Ackermann} \emph {et~al.}}]{Ackermann:2015zua}%
  \BibitemOpen
  \bibfield  {author} {\bibinfo {author} {\bibfnamefont {M.}~\bibnamefont
  {Ackermann}} \emph {et~al.} (\bibinfo {collaboration} {Fermi-LAT}),\ }\href
  {\doibase 10.1103/PhysRevLett.115.231301} {\bibfield  {journal} {\bibinfo
  {journal} {Phys. Rev. Lett.}\ }\textbf {\bibinfo {volume} {115}},\ \bibinfo
  {pages} {231301} (\bibinfo {year} {2015})},\ \Eprint
  {http://arxiv.org/abs/1503.02641} {arXiv:1503.02641 [astro-ph.HE]}
  \BibitemShut {NoStop}%
%%CITATION = ARXIV:1503.02641;%%
\bibitem [{\citenamefont {Ahnen}\ \emph {et~al.}(2016)\citenamefont {Ahnen}
  \emph {et~al.}}]{Ahnen:2016qkx}%
  \BibitemOpen
  \bibfield  {author} {\bibinfo {author} {\bibfnamefont {M.~L.}\ \bibnamefont
  {Ahnen}} \emph {et~al.} (\bibinfo {collaboration} {Fermi-LAT, MAGIC}),\
  }\href {\doibase 10.1088/1475-7516/2016/02/039} {\bibfield  {journal}
  {\bibinfo  {journal} {JCAP}\ }\textbf {\bibinfo {volume} {1602}},\ \bibinfo
  {pages} {039} (\bibinfo {year} {2016})},\ \Eprint
  {http://arxiv.org/abs/1601.06590} {arXiv:1601.06590 [astro-ph.HE]}
  \BibitemShut {NoStop}%
%%CITATION = ARXIV:1601.06590;%%
\bibitem [{\citenamefont {Profumo}(2013)}]{Profumo:2013yn}%
  \BibitemOpen
  \bibfield  {author} {\bibinfo {author} {\bibfnamefont {S.}~\bibnamefont
  {Profumo}},\ }in\ \href {\doibase 10.1142/9789814525220_0004} {\emph
  {\bibinfo {booktitle} {{Proceedings, Theoretical Advanced Study Institute in
  Elementary Particle Physics: Searching for New Physics at Small and Large
  Scales (TASI 2012): Boulder, Colorado, June 4-29, 2012}}}}\ (\bibinfo {year}
  {2013})\ pp.\ \bibinfo {pages} {143--189},\ \Eprint
  {http://arxiv.org/abs/1301.0952} {arXiv:1301.0952 [hep-ph]} \BibitemShut
  {NoStop}%
%%CITATION = ARXIV:1301.0952;%%
\bibitem [{\citenamefont {Lopez}\ \emph {et~al.}(2016)\citenamefont {Lopez},
  \citenamefont {Savage}, \citenamefont {Spolyar},\ and\ \citenamefont
  {Adams}}]{Lopez:2015uma}%
  \BibitemOpen
  \bibfield  {author} {\bibinfo {author} {\bibfnamefont {A.}~\bibnamefont
  {Lopez}}, \bibinfo {author} {\bibfnamefont {C.}~\bibnamefont {Savage}},
  \bibinfo {author} {\bibfnamefont {D.}~\bibnamefont {Spolyar}}, \ and\
  \bibinfo {author} {\bibfnamefont {D.~Q.}\ \bibnamefont {Adams}},\ }\bibfield
  {booktitle} {\emph {\bibinfo {booktitle} {{Proceedings, Meeting of the APS
  Division of Particles and Fields (DPF 2015): Ann Arbor, Michigan, USA, 4-8
  Aug 2015}}},\ }\href {\doibase 10.1088/1475-7516/2016/03/033} {\bibfield
  {journal} {\bibinfo  {journal} {JCAP}\ }\textbf {\bibinfo {volume} {1603}},\
  \bibinfo {pages} {033} (\bibinfo {year} {2016})},\ \Eprint
  {http://arxiv.org/abs/1501.01618} {arXiv:1501.01618 [astro-ph.CO]}
  \BibitemShut {NoStop}%
%%CITATION = ARXIV:1501.01618;%%
\bibitem [{\citenamefont {Tavani}\ \emph {et~al.}(2017)\citenamefont {Tavani}
  \emph {et~al.}}]{DeAngelis:2017gra}%
  \BibitemOpen
  \bibfield  {author} {\bibinfo {author} {\bibfnamefont {M.}~\bibnamefont
  {Tavani}} \emph {et~al.} (\bibinfo {collaboration} {e-ASTROGAM}),\
  }\href@noop {} {\  (\bibinfo {year} {2017})},\ \Eprint
  {http://arxiv.org/abs/1711.01265} {arXiv:1711.01265 [astro-ph.HE]}
  \BibitemShut {NoStop}%
%%CITATION = ARXIV:1711.01265;%%
\bibitem [{\citenamefont {Acharya}\ \emph {et~al.}(2017)\citenamefont {Acharya}
  \emph {et~al.}}]{Acharya:2017ttl}%
  \BibitemOpen
  \bibfield  {author} {\bibinfo {author} {\bibfnamefont {B.~S.}\ \bibnamefont
  {Acharya}} \emph {et~al.} (\bibinfo {collaboration} {Cherenkov Telescope
  Array Consortium}),\ }\href@noop {} {\  (\bibinfo {year} {2017})},\ \Eprint
  {http://arxiv.org/abs/1709.07997} {arXiv:1709.07997 [astro-ph.IM]}
  \BibitemShut {NoStop}%
%%CITATION = ARXIV:1709.07997;%%
\bibitem [{\citenamefont {Slatyer}(2016)}]{Slatyer:2015kla}%
  \BibitemOpen
  \bibfield  {author} {\bibinfo {author} {\bibfnamefont {T.~R.}\ \bibnamefont
  {Slatyer}},\ }\href {\doibase 10.1103/PhysRevD.93.023521} {\bibfield
  {journal} {\bibinfo  {journal} {Phys. Rev.}\ }\textbf {\bibinfo {volume}
  {D93}},\ \bibinfo {pages} {023521} (\bibinfo {year} {2016})},\ \Eprint
  {http://arxiv.org/abs/1506.03812} {arXiv:1506.03812 [astro-ph.CO]}
  \BibitemShut {NoStop}%
%%CITATION = ARXIV:1506.03812;%%
\bibitem [{\citenamefont {Slatyer}\ and\ \citenamefont
  {Wu}(2017)}]{Slatyer:2016qyl}%
  \BibitemOpen
  \bibfield  {author} {\bibinfo {author} {\bibfnamefont {T.~R.}\ \bibnamefont
  {Slatyer}}\ and\ \bibinfo {author} {\bibfnamefont {C.-L.}\ \bibnamefont
  {Wu}},\ }\href {\doibase 10.1103/PhysRevD.95.023010} {\bibfield  {journal}
  {\bibinfo  {journal} {Phys. Rev.}\ }\textbf {\bibinfo {volume} {D95}},\
  \bibinfo {pages} {023010} (\bibinfo {year} {2017})},\ \Eprint
  {http://arxiv.org/abs/1610.06933} {arXiv:1610.06933 [astro-ph.CO]}
  \BibitemShut {NoStop}%
%%CITATION = ARXIV:1610.06933;%%
\bibitem [{\citenamefont {Liu}\ \emph {et~al.}(2016)\citenamefont {Liu},
  \citenamefont {Slatyer},\ and\ \citenamefont {Zavala}}]{Liu:2016cnk}%
  \BibitemOpen
  \bibfield  {author} {\bibinfo {author} {\bibfnamefont {H.}~\bibnamefont
  {Liu}}, \bibinfo {author} {\bibfnamefont {T.~R.}\ \bibnamefont {Slatyer}}, \
  and\ \bibinfo {author} {\bibfnamefont {J.}~\bibnamefont {Zavala}},\ }\href
  {\doibase 10.1103/PhysRevD.94.063507} {\bibfield  {journal} {\bibinfo
  {journal} {Phys. Rev.}\ }\textbf {\bibinfo {volume} {D94}},\ \bibinfo {pages}
  {063507} (\bibinfo {year} {2016})},\ \Eprint
  {http://arxiv.org/abs/1604.02457} {arXiv:1604.02457 [astro-ph.CO]}
  \BibitemShut {NoStop}%
%%CITATION = ARXIV:1604.02457;%%
\bibitem [{\citenamefont {Bernal}\ \emph {et~al.}(2016)\citenamefont {Bernal},
  \citenamefont {Necib},\ and\ \citenamefont {Slatyer}}]{Bernal:2016guq}%
  \BibitemOpen
  \bibfield  {author} {\bibinfo {author} {\bibfnamefont {N.}~\bibnamefont
  {Bernal}}, \bibinfo {author} {\bibfnamefont {L.}~\bibnamefont {Necib}}, \
  and\ \bibinfo {author} {\bibfnamefont {T.~R.}\ \bibnamefont {Slatyer}},\
  }\href {\doibase 10.1088/1475-7516/2016/12/030} {\bibfield  {journal}
  {\bibinfo  {journal} {JCAP}\ }\textbf {\bibinfo {volume} {1612}},\ \bibinfo
  {pages} {030} (\bibinfo {year} {2016})},\ \Eprint
  {http://arxiv.org/abs/1606.00433} {arXiv:1606.00433 [astro-ph.CO]}
  \BibitemShut {NoStop}%
%%CITATION = ARXIV:1606.00433;%%
\bibitem [{\citenamefont {Dutra}\ \emph {et~al.}(2018)\citenamefont {Dutra},
  \citenamefont {Lindner}, \citenamefont {Profumo}, \citenamefont {Queiroz},
  \citenamefont {Rodejohann},\ and\ \citenamefont {Siqueira}}]{Dutra:2018gmv}%
  \BibitemOpen
  \bibfield  {author} {\bibinfo {author} {\bibfnamefont {M.}~\bibnamefont
  {Dutra}}, \bibinfo {author} {\bibfnamefont {M.}~\bibnamefont {Lindner}},
  \bibinfo {author} {\bibfnamefont {S.}~\bibnamefont {Profumo}}, \bibinfo
  {author} {\bibfnamefont {F.~S.}\ \bibnamefont {Queiroz}}, \bibinfo {author}
  {\bibfnamefont {W.}~\bibnamefont {Rodejohann}}, \ and\ \bibinfo {author}
  {\bibfnamefont {C.}~\bibnamefont {Siqueira}},\ }\href {\doibase
  10.1088/1475-7516/2018/03/037} {\bibfield  {journal} {\bibinfo  {journal}
  {JCAP}\ }\textbf {\bibinfo {volume} {1803}},\ \bibinfo {pages} {037}
  (\bibinfo {year} {2018})},\ \Eprint {http://arxiv.org/abs/1801.05447}
  {arXiv:1801.05447 [hep-ph]} \BibitemShut {NoStop}%
%%CITATION = ARXIV:1801.05447;%%
\bibitem [{\citenamefont {Goodman}\ \emph {et~al.}(2010)\citenamefont
  {Goodman}, \citenamefont {Ibe}, \citenamefont {Rajaraman}, \citenamefont
  {Shepherd}, \citenamefont {Tait} \emph {et~al.}}]{Goodman:2010ku}%
  \BibitemOpen
  \bibfield  {author} {\bibinfo {author} {\bibfnamefont {J.}~\bibnamefont
  {Goodman}}, \bibinfo {author} {\bibfnamefont {M.}~\bibnamefont {Ibe}},
  \bibinfo {author} {\bibfnamefont {A.}~\bibnamefont {Rajaraman}}, \bibinfo
  {author} {\bibfnamefont {W.}~\bibnamefont {Shepherd}}, \bibinfo {author}
  {\bibfnamefont {T.~M.}\ \bibnamefont {Tait}},  \emph {et~al.},\ }\href
  {\doibase 10.1103/PhysRevD.82.116010} {\bibfield  {journal} {\bibinfo
  {journal} {Phys.Rev.}\ }\textbf {\bibinfo {volume} {D82}},\ \bibinfo {pages}
  {116010} (\bibinfo {year} {2010})},\ \Eprint {http://arxiv.org/abs/1008.1783}
  {arXiv:1008.1783 [hep-ph]} \BibitemShut {NoStop}%
%%CITATION = ARXIV:1008.1783;%%
\bibitem [{\citenamefont {Rajaraman}\ \emph {et~al.}(2011)\citenamefont
  {Rajaraman}, \citenamefont {Shepherd}, \citenamefont {Tait},\ and\
  \citenamefont {Wijangco}}]{Rajaraman:2011wf}%
  \BibitemOpen
  \bibfield  {author} {\bibinfo {author} {\bibfnamefont {A.}~\bibnamefont
  {Rajaraman}}, \bibinfo {author} {\bibfnamefont {W.}~\bibnamefont {Shepherd}},
  \bibinfo {author} {\bibfnamefont {T.~M.~P.}\ \bibnamefont {Tait}}, \ and\
  \bibinfo {author} {\bibfnamefont {A.~M.}\ \bibnamefont {Wijangco}},\ }\href
  {\doibase 10.1103/PhysRevD.84.095013} {\bibfield  {journal} {\bibinfo
  {journal} {Phys. Rev.}\ }\textbf {\bibinfo {volume} {D84}},\ \bibinfo {pages}
  {095013} (\bibinfo {year} {2011})},\ \Eprint {http://arxiv.org/abs/1108.1196}
  {arXiv:1108.1196 [hep-ph]} \BibitemShut {NoStop}%
%%CITATION = ARXIV:1108.1196;%%
\bibitem [{\citenamefont {Bauer}\ \emph {et~al.}(2015)\citenamefont {Bauer}
  \emph {et~al.}}]{Bauer:2013ihz}%
  \BibitemOpen
  \bibfield  {author} {\bibinfo {author} {\bibfnamefont {D.}~\bibnamefont
  {Bauer}} \emph {et~al.} (\bibinfo {collaboration} {Snowmass 2013 Cosmic
  Frontier Working Groups 1–4}),\ }\href {\doibase
  10.1016/j.dark.2015.04.001} {\bibfield  {journal} {\bibinfo  {journal} {Phys.
  Dark Univ.}\ }\textbf {\bibinfo {volume} {7-8}},\ \bibinfo {pages} {16}
  (\bibinfo {year} {2015})},\ \Eprint {http://arxiv.org/abs/1305.1605}
  {arXiv:1305.1605 [hep-ph]} \BibitemShut {NoStop}%
%%CITATION = ARXIV:1305.1605;%%
\bibitem [{\citenamefont {Szczerbinska}\ \emph {et~al.}(2016)\citenamefont
  {Szczerbinska}, \citenamefont {Allahverdi}, \citenamefont {Babu},
  \citenamefont {Balantekin}, \citenamefont {Dutta}, \citenamefont {Kamon},
  \citenamefont {Kumar}, \citenamefont {Queiroz}, \citenamefont {Strigari},\
  and\ \citenamefont {Surman}}]{Proceedings:2016rni}%
  \BibitemOpen
  \bibinfo {editor} {\bibfnamefont {B.}~\bibnamefont {Szczerbinska}}, \bibinfo
  {editor} {\bibfnamefont {R.}~\bibnamefont {Allahverdi}}, \bibinfo {editor}
  {\bibfnamefont {K.}~\bibnamefont {Babu}}, \bibinfo {editor} {\bibfnamefont
  {B.}~\bibnamefont {Balantekin}}, \bibinfo {editor} {\bibfnamefont
  {B.}~\bibnamefont {Dutta}}, \bibinfo {editor} {\bibfnamefont
  {T.}~\bibnamefont {Kamon}}, \bibinfo {editor} {\bibfnamefont
  {J.}~\bibnamefont {Kumar}}, \bibinfo {editor} {\bibfnamefont
  {F.}~\bibnamefont {Queiroz}}, \bibinfo {editor} {\bibfnamefont
  {L.}~\bibnamefont {Strigari}}, \ and\ \bibinfo {editor} {\bibfnamefont
  {R.}~\bibnamefont {Surman}},\ eds.,\ \href
  {http://aip.scitation.org/toc/apc/1743/1} {\emph {\bibinfo {title}
  {{Proceedings, Workshop on Neutrino Physics : Session of CETUP* 2015 and 9th
  International Conference on Interconnections between Particle Physics and
  Cosmology (PPC2015)}}}},\ Vol.\ \bibinfo {volume} {1743}\ (\bibinfo {year}
  {2016})\BibitemShut {NoStop}%
%%CITATION = APCPC,1743,;%%
\bibitem [{\citenamefont {Aad}\ \emph {et~al.}(2014)\citenamefont {Aad} \emph
  {et~al.}}]{Aad:2014cka}%
  \BibitemOpen
  \bibfield  {author} {\bibinfo {author} {\bibfnamefont {G.}~\bibnamefont
  {Aad}} \emph {et~al.} (\bibinfo {collaboration} {ATLAS}),\ }\href {\doibase
  10.1103/PhysRevD.90.052005} {\bibfield  {journal} {\bibinfo  {journal} {Phys.
  Rev.}\ }\textbf {\bibinfo {volume} {D90}},\ \bibinfo {pages} {052005}
  (\bibinfo {year} {2014})},\ \Eprint {http://arxiv.org/abs/1405.4123}
  {arXiv:1405.4123 [hep-ex]} \BibitemShut {NoStop}%
%%CITATION = ARXIV:1405.4123;%%
\bibitem [{\citenamefont {Khachatryan}\ \emph {et~al.}(2017)\citenamefont
  {Khachatryan} \emph {et~al.}}]{Khachatryan:2016zqb}%
  \BibitemOpen
  \bibfield  {author} {\bibinfo {author} {\bibfnamefont {V.}~\bibnamefont
  {Khachatryan}} \emph {et~al.} (\bibinfo {collaboration} {CMS}),\ }\href
  {\doibase 10.1016/j.physletb.2017.02.010} {\bibfield  {journal} {\bibinfo
  {journal} {Phys. Lett.}\ }\textbf {\bibinfo {volume} {B768}},\ \bibinfo
  {pages} {57} (\bibinfo {year} {2017})},\ \Eprint
  {http://arxiv.org/abs/1609.05391} {arXiv:1609.05391 [hep-ex]} \BibitemShut
  {NoStop}%
%%CITATION = ARXIV:1609.05391;%%
\bibitem [{\citenamefont {Sirunyan}\ \emph {et~al.}(2018)\citenamefont
  {Sirunyan} \emph {et~al.}}]{Sirunyan:2018exx}%
  \BibitemOpen
  \bibfield  {author} {\bibinfo {author} {\bibfnamefont {A.~M.}\ \bibnamefont
  {Sirunyan}} \emph {et~al.} (\bibinfo {collaboration} {CMS}),\ }\href@noop {}
  {\  (\bibinfo {year} {2018})},\ \Eprint {http://arxiv.org/abs/1803.06292}
  {arXiv:1803.06292 [hep-ex]} \BibitemShut {NoStop}%
%%CITATION = ARXIV:1803.06292;%%
\bibitem [{\citenamefont {Lopez}\ \emph {et~al.}(1992)\citenamefont {Lopez},
  \citenamefont {Nanopoulos},\ and\ \citenamefont {Yuan}}]{Lopez:1991pp}%
  \BibitemOpen
  \bibfield  {author} {\bibinfo {author} {\bibfnamefont {J.~L.}\ \bibnamefont
  {Lopez}}, \bibinfo {author} {\bibfnamefont {D.~V.}\ \bibnamefont
  {Nanopoulos}}, \ and\ \bibinfo {author} {\bibfnamefont {K.-j.}\ \bibnamefont
  {Yuan}},\ }\href {\doibase 10.1016/0550-3213(92)90293-K} {\bibfield
  {journal} {\bibinfo  {journal} {Nucl. Phys.}\ }\textbf {\bibinfo {volume}
  {B370}},\ \bibinfo {pages} {445} (\bibinfo {year} {1992})}\BibitemShut
  {NoStop}%
%%CITATION = NUPHA,B370,445;%%
\bibitem [{\citenamefont {Baer}\ \emph {et~al.}(2008)\citenamefont {Baer},
  \citenamefont {Kraml}, \citenamefont {Sekmen},\ and\ \citenamefont
  {Summy}}]{Baer:2008jn}%
  \BibitemOpen
  \bibfield  {author} {\bibinfo {author} {\bibfnamefont {H.}~\bibnamefont
  {Baer}}, \bibinfo {author} {\bibfnamefont {S.}~\bibnamefont {Kraml}},
  \bibinfo {author} {\bibfnamefont {S.}~\bibnamefont {Sekmen}}, \ and\ \bibinfo
  {author} {\bibfnamefont {H.}~\bibnamefont {Summy}},\ }\href {\doibase
  10.1088/1126-6708/2008/03/056} {\bibfield  {journal} {\bibinfo  {journal}
  {JHEP}\ }\textbf {\bibinfo {volume} {03}},\ \bibinfo {pages} {056} (\bibinfo
  {year} {2008})},\ \Eprint {http://arxiv.org/abs/0801.1831} {arXiv:0801.1831
  [hep-ph]} \BibitemShut {NoStop}%
%%CITATION = ARXIV:0801.1831;%%
\bibitem [{\citenamefont {Baer}\ and\ \citenamefont
  {Summy}(2008)}]{Baer:2008eq}%
  \BibitemOpen
  \bibfield  {author} {\bibinfo {author} {\bibfnamefont {H.}~\bibnamefont
  {Baer}}\ and\ \bibinfo {author} {\bibfnamefont {H.}~\bibnamefont {Summy}},\
  }\href {\doibase 10.1016/j.physletb.2008.06.072} {\bibfield  {journal}
  {\bibinfo  {journal} {Phys. Lett.}\ }\textbf {\bibinfo {volume} {B666}},\
  \bibinfo {pages} {5} (\bibinfo {year} {2008})},\ \Eprint
  {http://arxiv.org/abs/0803.0510} {arXiv:0803.0510 [hep-ph]} \BibitemShut
  {NoStop}%
%%CITATION = ARXIV:0803.0510;%%
\bibitem [{\citenamefont {Agashe}\ \emph {et~al.}(2010)\citenamefont {Agashe},
  \citenamefont {Blum}, \citenamefont {Lee},\ and\ \citenamefont
  {Perez}}]{Agashe:2009ja}%
  \BibitemOpen
  \bibfield  {author} {\bibinfo {author} {\bibfnamefont {K.}~\bibnamefont
  {Agashe}}, \bibinfo {author} {\bibfnamefont {K.}~\bibnamefont {Blum}},
  \bibinfo {author} {\bibfnamefont {S.~J.}\ \bibnamefont {Lee}}, \ and\
  \bibinfo {author} {\bibfnamefont {G.}~\bibnamefont {Perez}},\ }\href
  {\doibase 10.1103/PhysRevD.81.075012} {\bibfield  {journal} {\bibinfo
  {journal} {Phys. Rev.}\ }\textbf {\bibinfo {volume} {D81}},\ \bibinfo {pages}
  {075012} (\bibinfo {year} {2010})},\ \Eprint {http://arxiv.org/abs/0912.3070}
  {arXiv:0912.3070 [hep-ph]} \BibitemShut {NoStop}%
%%CITATION = ARXIV:0912.3070;%%
\bibitem [{\citenamefont {Chakrabortty}\ \emph {et~al.}(2014)\citenamefont
  {Chakrabortty}, \citenamefont {Mohanty},\ and\ \citenamefont
  {Rao}}]{Chakrabortty:2013voa}%
  \BibitemOpen
  \bibfield  {author} {\bibinfo {author} {\bibfnamefont {J.}~\bibnamefont
  {Chakrabortty}}, \bibinfo {author} {\bibfnamefont {S.}~\bibnamefont
  {Mohanty}}, \ and\ \bibinfo {author} {\bibfnamefont {S.}~\bibnamefont
  {Rao}},\ }\href {\doibase 10.1007/JHEP02(2014)074} {\bibfield  {journal}
  {\bibinfo  {journal} {JHEP}\ }\textbf {\bibinfo {volume} {02}},\ \bibinfo
  {pages} {074} (\bibinfo {year} {2014})},\ \Eprint
  {http://arxiv.org/abs/1310.3620} {arXiv:1310.3620 [hep-ph]} \BibitemShut
  {NoStop}%
%%CITATION = ARXIV:1310.3620;%%
\bibitem [{\citenamefont {Anandakrishnan}\ and\ \citenamefont
  {Sinha}(2014)}]{Anandakrishnan:2013tqa}%
  \BibitemOpen
  \bibfield  {author} {\bibinfo {author} {\bibfnamefont {A.}~\bibnamefont
  {Anandakrishnan}}\ and\ \bibinfo {author} {\bibfnamefont {K.}~\bibnamefont
  {Sinha}},\ }\href {\doibase 10.1103/PhysRevD.89.055015} {\bibfield  {journal}
  {\bibinfo  {journal} {Phys. Rev.}\ }\textbf {\bibinfo {volume} {D89}},\
  \bibinfo {pages} {055015} (\bibinfo {year} {2014})},\ \Eprint
  {http://arxiv.org/abs/1310.7579} {arXiv:1310.7579 [hep-ph]} \BibitemShut
  {NoStop}%
%%CITATION = ARXIV:1310.7579;%%
\bibitem [{\citenamefont {Ko}\ \emph {et~al.}(2015)\citenamefont {Ko},
  \citenamefont {Omura},\ and\ \citenamefont {Yu}}]{Ko:2015fxa}%
  \BibitemOpen
  \bibfield  {author} {\bibinfo {author} {\bibfnamefont {P.}~\bibnamefont
  {Ko}}, \bibinfo {author} {\bibfnamefont {Y.}~\bibnamefont {Omura}}, \ and\
  \bibinfo {author} {\bibfnamefont {C.}~\bibnamefont {Yu}},\ }\href {\doibase
  10.1007/JHEP06(2015)034} {\bibfield  {journal} {\bibinfo  {journal} {JHEP}\
  }\textbf {\bibinfo {volume} {06}},\ \bibinfo {pages} {034} (\bibinfo {year}
  {2015})},\ \Eprint {http://arxiv.org/abs/1502.00262} {arXiv:1502.00262
  [hep-ph]} \BibitemShut {NoStop}%
%%CITATION = ARXIV:1502.00262;%%
\bibitem [{\citenamefont {Arbelaez}\ \emph {et~al.}(2016)\citenamefont
  {Arbelaez}, \citenamefont {Longas}, \citenamefont {Restrepo},\ and\
  \citenamefont {Zapata}}]{Arbelaez:2015ila}%
  \BibitemOpen
  \bibfield  {author} {\bibinfo {author} {\bibfnamefont {C.}~\bibnamefont
  {Arbelaez}}, \bibinfo {author} {\bibfnamefont {R.}~\bibnamefont {Longas}},
  \bibinfo {author} {\bibfnamefont {D.}~\bibnamefont {Restrepo}}, \ and\
  \bibinfo {author} {\bibfnamefont {O.}~\bibnamefont {Zapata}},\ }\href
  {\doibase 10.1103/PhysRevD.93.013012} {\bibfield  {journal} {\bibinfo
  {journal} {Phys. Rev.}\ }\textbf {\bibinfo {volume} {D93}},\ \bibinfo {pages}
  {013012} (\bibinfo {year} {2016})},\ \Eprint
  {http://arxiv.org/abs/1509.06313} {arXiv:1509.06313 [hep-ph]} \BibitemShut
  {NoStop}%
%%CITATION = ARXIV:1509.06313;%%
\bibitem [{\citenamefont {Chen}\ \emph {et~al.}(2018)\citenamefont {Chen},
  \citenamefont {Gogoladze}, \citenamefont {Hu}, \citenamefont {Li},\ and\
  \citenamefont {Wu}}]{Chen:2017rpn}%
  \BibitemOpen
  \bibfield  {author} {\bibinfo {author} {\bibfnamefont {H.-Y.}\ \bibnamefont
  {Chen}}, \bibinfo {author} {\bibfnamefont {I.}~\bibnamefont {Gogoladze}},
  \bibinfo {author} {\bibfnamefont {S.}~\bibnamefont {Hu}}, \bibinfo {author}
  {\bibfnamefont {T.}~\bibnamefont {Li}}, \ and\ \bibinfo {author}
  {\bibfnamefont {L.}~\bibnamefont {Wu}},\ }\href {\doibase
  10.1140/epjc/s10052-017-5496-z} {\bibfield  {journal} {\bibinfo  {journal}
  {Eur. Phys. J.}\ }\textbf {\bibinfo {volume} {C78}},\ \bibinfo {pages} {26}
  (\bibinfo {year} {2018})},\ \Eprint {http://arxiv.org/abs/1703.07542}
  {arXiv:1703.07542 [hep-ph]} \BibitemShut {NoStop}%
%%CITATION = ARXIV:1703.07542;%%
\bibitem [{\citenamefont {Arcadi}\ \emph
  {et~al.}(2017{\natexlab{b}})\citenamefont {Arcadi}, \citenamefont {Lindner},
  \citenamefont {Mambrini}, \citenamefont {Pierre},\ and\ \citenamefont
  {Queiroz}}]{Arcadi:2017atc}%
  \BibitemOpen
  \bibfield  {author} {\bibinfo {author} {\bibfnamefont {G.}~\bibnamefont
  {Arcadi}}, \bibinfo {author} {\bibfnamefont {M.}~\bibnamefont {Lindner}},
  \bibinfo {author} {\bibfnamefont {Y.}~\bibnamefont {Mambrini}}, \bibinfo
  {author} {\bibfnamefont {M.}~\bibnamefont {Pierre}}, \ and\ \bibinfo {author}
  {\bibfnamefont {F.~S.}\ \bibnamefont {Queiroz}},\ }\href {\doibase
  10.1016/j.physletb.2017.05.023} {\bibfield  {journal} {\bibinfo  {journal}
  {Phys. Lett.}\ }\textbf {\bibinfo {volume} {B771}},\ \bibinfo {pages} {508}
  (\bibinfo {year} {2017}{\natexlab{b}})},\ \Eprint
  {http://arxiv.org/abs/1704.02328} {arXiv:1704.02328 [hep-ph]} \BibitemShut
  {NoStop}%
%%CITATION = ARXIV:1704.02328;%%
\bibitem [{\citenamefont {Coriano}\ and\ \citenamefont
  {Frampton}(2017)}]{Coriano:2017ghp}%
  \BibitemOpen
  \bibfield  {author} {\bibinfo {author} {\bibfnamefont {C.}~\bibnamefont
  {Coriano}}\ and\ \bibinfo {author} {\bibfnamefont {P.~H.}\ \bibnamefont
  {Frampton}},\ }\href@noop {} {\  (\bibinfo {year} {2017})},\ \Eprint
  {http://arxiv.org/abs/1712.03865} {arXiv:1712.03865 [hep-ph]} \BibitemShut
  {NoStop}%
%%CITATION = ARXIV:1712.03865;%%
\bibitem [{\citenamefont {King}\ \emph {et~al.}(2017)\citenamefont {King},
  \citenamefont {King},\ and\ \citenamefont {Moretti}}]{King:2017cwv}%
  \BibitemOpen
  \bibfield  {author} {\bibinfo {author} {\bibfnamefont {S.~J.~D.}\
  \bibnamefont {King}}, \bibinfo {author} {\bibfnamefont {S.~F.}\ \bibnamefont
  {King}}, \ and\ \bibinfo {author} {\bibfnamefont {S.}~\bibnamefont
  {Moretti}},\ }\href@noop {} {\  (\bibinfo {year} {2017})},\ \Eprint
  {http://arxiv.org/abs/1712.01279} {arXiv:1712.01279 [hep-ph]} \BibitemShut
  {NoStop}%
%%CITATION = ARXIV:1712.01279;%%
\bibitem [{\citenamefont {Patrignani}\ \emph {et~al.}(2016)\citenamefont
  {Patrignani} \emph {et~al.}}]{Patrignani:2016xqp}%
  \BibitemOpen
  \bibfield  {author} {\bibinfo {author} {\bibfnamefont {C.}~\bibnamefont
  {Patrignani}} \emph {et~al.} (\bibinfo {collaboration} {Particle Data
  Group}),\ }\href {\doibase 10.1088/1674-1137/40/10/100001} {\bibfield
  {journal} {\bibinfo  {journal} {Chin. Phys.}\ }\textbf {\bibinfo {volume}
  {C40}},\ \bibinfo {pages} {100001} (\bibinfo {year} {2016})}\BibitemShut
  {NoStop}%
%%CITATION = CHPHD,C40,100001;%%
\bibitem [{\citenamefont {Bell}\ \emph
  {et~al.}(2017{\natexlab{a}})\citenamefont {Bell}, \citenamefont {Busoni},\
  and\ \citenamefont {Sanderson}}]{Bell:2016ekl}%
  \BibitemOpen
  \bibfield  {author} {\bibinfo {author} {\bibfnamefont {N.~F.}\ \bibnamefont
  {Bell}}, \bibinfo {author} {\bibfnamefont {G.}~\bibnamefont {Busoni}}, \ and\
  \bibinfo {author} {\bibfnamefont {I.~W.}\ \bibnamefont {Sanderson}},\ }\href
  {\doibase 10.1088/1475-7516/2017/03/015} {\bibfield  {journal} {\bibinfo
  {journal} {JCAP}\ }\textbf {\bibinfo {volume} {1703}},\ \bibinfo {pages}
  {015} (\bibinfo {year} {2017}{\natexlab{a}})},\ \Eprint
  {http://arxiv.org/abs/1612.03475} {arXiv:1612.03475 [hep-ph]} \BibitemShut
  {NoStop}%
%%CITATION = ARXIV:1612.03475;%%
\bibitem [{\citenamefont {Bell}\ \emph {et~al.}(2016)\citenamefont {Bell},
  \citenamefont {Busoni}, \citenamefont {Kobakhidze}, \citenamefont {Long},\
  and\ \citenamefont {Schmidt}}]{Bell:2016obu}%
  \BibitemOpen
  \bibfield  {author} {\bibinfo {author} {\bibfnamefont {N.}~\bibnamefont
  {Bell}}, \bibinfo {author} {\bibfnamefont {G.}~\bibnamefont {Busoni}},
  \bibinfo {author} {\bibfnamefont {A.}~\bibnamefont {Kobakhidze}}, \bibinfo
  {author} {\bibfnamefont {D.~M.}\ \bibnamefont {Long}}, \ and\ \bibinfo
  {author} {\bibfnamefont {M.~A.}\ \bibnamefont {Schmidt}},\ }\href {\doibase
  10.1007/JHEP08(2016)125} {\bibfield  {journal} {\bibinfo  {journal} {JHEP}\
  }\textbf {\bibinfo {volume} {08}},\ \bibinfo {pages} {125} (\bibinfo {year}
  {2016})},\ \Eprint {http://arxiv.org/abs/1606.02722} {arXiv:1606.02722
  [hep-ph]} \BibitemShut {NoStop}%
%%CITATION = ARXIV:1606.02722;%%
\bibitem [{\citenamefont {Ismail}\ \emph
  {et~al.}(2017{\natexlab{a}})\citenamefont {Ismail}, \citenamefont {Keung},
  \citenamefont {Tsao},\ and\ \citenamefont {Unwin}}]{Ismail:2016tod}%
  \BibitemOpen
  \bibfield  {author} {\bibinfo {author} {\bibfnamefont {A.}~\bibnamefont
  {Ismail}}, \bibinfo {author} {\bibfnamefont {W.-Y.}\ \bibnamefont {Keung}},
  \bibinfo {author} {\bibfnamefont {K.-H.}\ \bibnamefont {Tsao}}, \ and\
  \bibinfo {author} {\bibfnamefont {J.}~\bibnamefont {Unwin}},\ }\href
  {\doibase 10.1016/j.nuclphysb.2017.03.001} {\bibfield  {journal} {\bibinfo
  {journal} {Nucl. Phys.}\ }\textbf {\bibinfo {volume} {B918}},\ \bibinfo
  {pages} {220} (\bibinfo {year} {2017}{\natexlab{a}})},\ \Eprint
  {http://arxiv.org/abs/1609.02188} {arXiv:1609.02188 [hep-ph]} \BibitemShut
  {NoStop}%
%%CITATION = ARXIV:1609.02188;%%
\bibitem [{\citenamefont {Ismail}\ \emph
  {et~al.}(2017{\natexlab{b}})\citenamefont {Ismail}, \citenamefont {Katz},\
  and\ \citenamefont {Racco}}]{Ismail:2017ulg}%
  \BibitemOpen
  \bibfield  {author} {\bibinfo {author} {\bibfnamefont {A.}~\bibnamefont
  {Ismail}}, \bibinfo {author} {\bibfnamefont {A.}~\bibnamefont {Katz}}, \ and\
  \bibinfo {author} {\bibfnamefont {D.}~\bibnamefont {Racco}},\ }\href
  {\doibase 10.1007/JHEP10(2017)165} {\bibfield  {journal} {\bibinfo  {journal}
  {JHEP}\ }\textbf {\bibinfo {volume} {10}},\ \bibinfo {pages} {165} (\bibinfo
  {year} {2017}{\natexlab{b}})},\ \Eprint {http://arxiv.org/abs/1707.00709}
  {arXiv:1707.00709 [hep-ph]} \BibitemShut {NoStop}%
%%CITATION = ARXIV:1707.00709;%%
\bibitem [{\citenamefont {Cho}\ \emph {et~al.}(1998)\citenamefont {Cho},
  \citenamefont {Hagiwara},\ and\ \citenamefont {Umeda}}]{Cho:1998nr}%
  \BibitemOpen
  \bibfield  {author} {\bibinfo {author} {\bibfnamefont {G.-C.}\ \bibnamefont
  {Cho}}, \bibinfo {author} {\bibfnamefont {K.}~\bibnamefont {Hagiwara}}, \
  and\ \bibinfo {author} {\bibfnamefont {Y.}~\bibnamefont {Umeda}},\ }\href
  {\doibase 10.1016/S0550-3213(98)00549-5, 10.1016/S0550-3213(99)00382-X,
  10.1016/S0550-3213(99)00480-0} {\bibfield  {journal} {\bibinfo  {journal}
  {Nucl. Phys.}\ }\textbf {\bibinfo {volume} {B531}},\ \bibinfo {pages} {65}
  (\bibinfo {year} {1998})},\ \bibinfo {note} {[Erratum: Nucl.
  Phys.B565,483(2000)]},\ \Eprint {http://arxiv.org/abs/hep-ph/9805448}
  {arXiv:hep-ph/9805448 [hep-ph]} \BibitemShut {NoStop}%
%%CITATION = HEP-PH/9805448;%%
\bibitem [{\citenamefont {Nie}\ and\ \citenamefont {Sher}(2001)}]{Nie:2001ti}%
  \BibitemOpen
  \bibfield  {author} {\bibinfo {author} {\bibfnamefont {S.}~\bibnamefont
  {Nie}}\ and\ \bibinfo {author} {\bibfnamefont {M.}~\bibnamefont {Sher}},\
  }\href {\doibase 10.1103/PhysRevD.64.073015} {\bibfield  {journal} {\bibinfo
  {journal} {Phys. Rev.}\ }\textbf {\bibinfo {volume} {D64}},\ \bibinfo {pages}
  {073015} (\bibinfo {year} {2001})},\ \Eprint
  {http://arxiv.org/abs/hep-ph/0102139} {arXiv:hep-ph/0102139 [hep-ph]}
  \BibitemShut {NoStop}%
%%CITATION = HEP-PH/0102139;%%
\bibitem [{\citenamefont {Howl}\ and\ \citenamefont
  {King}(2008)}]{Howl:2007zi}%
  \BibitemOpen
  \bibfield  {author} {\bibinfo {author} {\bibfnamefont {R.}~\bibnamefont
  {Howl}}\ and\ \bibinfo {author} {\bibfnamefont {S.~F.}\ \bibnamefont
  {King}},\ }\href {\doibase 10.1088/1126-6708/2008/01/030} {\bibfield
  {journal} {\bibinfo  {journal} {JHEP}\ }\textbf {\bibinfo {volume} {01}},\
  \bibinfo {pages} {030} (\bibinfo {year} {2008})},\ \Eprint
  {http://arxiv.org/abs/0708.1451} {arXiv:0708.1451 [hep-ph]} \BibitemShut
  {NoStop}%
%%CITATION = ARXIV:0708.1451;%%
\bibitem [{\citenamefont {Stech}\ and\ \citenamefont
  {Tavartkiladze}(2008)}]{Stech:2008wd}%
  \BibitemOpen
  \bibfield  {author} {\bibinfo {author} {\bibfnamefont {B.}~\bibnamefont
  {Stech}}\ and\ \bibinfo {author} {\bibfnamefont {Z.}~\bibnamefont
  {Tavartkiladze}},\ }\href {\doibase 10.1103/PhysRevD.77.076009} {\bibfield
  {journal} {\bibinfo  {journal} {Phys. Rev.}\ }\textbf {\bibinfo {volume}
  {D77}},\ \bibinfo {pages} {076009} (\bibinfo {year} {2008})},\ \Eprint
  {http://arxiv.org/abs/0802.0894} {arXiv:0802.0894 [hep-ph]} \BibitemShut
  {NoStop}%
%%CITATION = ARXIV:0802.0894;%%
\bibitem [{\citenamefont {Rojas}\ and\ \citenamefont
  {Erler}(2015)}]{Rojas:2015tqa}%
  \BibitemOpen
  \bibfield  {author} {\bibinfo {author} {\bibfnamefont {E.}~\bibnamefont
  {Rojas}}\ and\ \bibinfo {author} {\bibfnamefont {J.}~\bibnamefont {Erler}},\
  }\href {\doibase 10.1007/JHEP10(2015)063} {\bibfield  {journal} {\bibinfo
  {journal} {JHEP}\ }\textbf {\bibinfo {volume} {10}},\ \bibinfo {pages} {063}
  (\bibinfo {year} {2015})},\ \Eprint {http://arxiv.org/abs/1505.03208}
  {arXiv:1505.03208 [hep-ph]} \BibitemShut {NoStop}%
%%CITATION = ARXIV:1505.03208;%%
\bibitem [{\citenamefont {Benli}\ and\ \citenamefont
  {Dereli}(2017)}]{Benli:2017eld}%
  \BibitemOpen
  \bibfield  {author} {\bibinfo {author} {\bibfnamefont {S.}~\bibnamefont
  {Benli}}\ and\ \bibinfo {author} {\bibfnamefont {T.}~\bibnamefont {Dereli}},\
  }\href {\doibase 10.1007/s10773-018-3757-8} {\  (\bibinfo {year} {2017}),\
  10.1007/s10773-018-3757-8},\ \Eprint {http://arxiv.org/abs/1707.03144}
  {arXiv:1707.03144 [hep-ph]} \BibitemShut {NoStop}%
%%CITATION = ARXIV:1707.03144;%%
\bibitem [{\citenamefont {Dutta}\ \emph {et~al.}(2018)\citenamefont {Dutta},
  \citenamefont {Ghosh}, \citenamefont {Gogoladze},\ and\ \citenamefont
  {Li}}]{Dutta:2018qei}%
  \BibitemOpen
  \bibfield  {author} {\bibinfo {author} {\bibfnamefont {B.}~\bibnamefont
  {Dutta}}, \bibinfo {author} {\bibfnamefont {S.}~\bibnamefont {Ghosh}},
  \bibinfo {author} {\bibfnamefont {I.}~\bibnamefont {Gogoladze}}, \ and\
  \bibinfo {author} {\bibfnamefont {T.}~\bibnamefont {Li}},\ }\href@noop {} {\
  (\bibinfo {year} {2018})},\ \Eprint {http://arxiv.org/abs/1805.01866}
  {arXiv:1805.01866 [hep-ph]} \BibitemShut {NoStop}%
%%CITATION = ARXIV:1805.01866;%%
\bibitem [{\citenamefont {Schwichtenberg}(2018)}]{Schwichtenberg:2017xhv}%
  \BibitemOpen
  \bibfield  {author} {\bibinfo {author} {\bibfnamefont {J.}~\bibnamefont
  {Schwichtenberg}},\ }\href {\doibase 10.1007/JHEP02(2018)016} {\bibfield
  {journal} {\bibinfo  {journal} {JHEP}\ }\textbf {\bibinfo {volume} {02}},\
  \bibinfo {pages} {016} (\bibinfo {year} {2018})},\ \Eprint
  {http://arxiv.org/abs/1704.04219} {arXiv:1704.04219 [hep-ph]} \BibitemShut
  {NoStop}%
%%CITATION = ARXIV:1704.04219;%%
\bibitem [{\citenamefont {Kahlhoefer}\ \emph {et~al.}(2016)\citenamefont
  {Kahlhoefer}, \citenamefont {Schmidt-Hoberg}, \citenamefont {Schwetz},\ and\
  \citenamefont {Vogl}}]{Kahlhoefer:2015bea}%
  \BibitemOpen
  \bibfield  {author} {\bibinfo {author} {\bibfnamefont {F.}~\bibnamefont
  {Kahlhoefer}}, \bibinfo {author} {\bibfnamefont {K.}~\bibnamefont
  {Schmidt-Hoberg}}, \bibinfo {author} {\bibfnamefont {T.}~\bibnamefont
  {Schwetz}}, \ and\ \bibinfo {author} {\bibfnamefont {S.}~\bibnamefont
  {Vogl}},\ }\href {\doibase 10.1007/JHEP02(2016)016} {\bibfield  {journal}
  {\bibinfo  {journal} {JHEP}\ }\textbf {\bibinfo {volume} {02}},\ \bibinfo
  {pages} {016} (\bibinfo {year} {2016})},\ \Eprint
  {http://arxiv.org/abs/1510.02110} {arXiv:1510.02110 [hep-ph]} \BibitemShut
  {NoStop}%
%%CITATION = ARXIV:1510.02110;%%
\bibitem [{\citenamefont {Celis}\ \emph {et~al.}(2017)\citenamefont {Celis},
  \citenamefont {Feng},\ and\ \citenamefont {Vollmann}}]{Celis:2016ayl}%
  \BibitemOpen
  \bibfield  {author} {\bibinfo {author} {\bibfnamefont {A.}~\bibnamefont
  {Celis}}, \bibinfo {author} {\bibfnamefont {W.-Z.}\ \bibnamefont {Feng}}, \
  and\ \bibinfo {author} {\bibfnamefont {M.}~\bibnamefont {Vollmann}},\ }\href
  {\doibase 10.1103/PhysRevD.95.035018} {\bibfield  {journal} {\bibinfo
  {journal} {Phys. Rev.}\ }\textbf {\bibinfo {volume} {D95}},\ \bibinfo {pages}
  {035018} (\bibinfo {year} {2017})},\ \Eprint
  {http://arxiv.org/abs/1608.03894} {arXiv:1608.03894 [hep-ph]} \BibitemShut
  {NoStop}%
%%CITATION = ARXIV:1608.03894;%%
\bibitem [{\citenamefont {Duerr}\ \emph {et~al.}(2016)\citenamefont {Duerr},
  \citenamefont {Kahlhoefer}, \citenamefont {Schmidt-Hoberg}, \citenamefont
  {Schwetz},\ and\ \citenamefont {Vogl}}]{Duerr:2016tmh}%
  \BibitemOpen
  \bibfield  {author} {\bibinfo {author} {\bibfnamefont {M.}~\bibnamefont
  {Duerr}}, \bibinfo {author} {\bibfnamefont {F.}~\bibnamefont {Kahlhoefer}},
  \bibinfo {author} {\bibfnamefont {K.}~\bibnamefont {Schmidt-Hoberg}},
  \bibinfo {author} {\bibfnamefont {T.}~\bibnamefont {Schwetz}}, \ and\
  \bibinfo {author} {\bibfnamefont {S.}~\bibnamefont {Vogl}},\ }\href {\doibase
  10.1007/JHEP09(2016)042} {\bibfield  {journal} {\bibinfo  {journal} {JHEP}\
  }\textbf {\bibinfo {volume} {09}},\ \bibinfo {pages} {042} (\bibinfo {year}
  {2016})},\ \Eprint {http://arxiv.org/abs/1606.07609} {arXiv:1606.07609
  [hep-ph]} \BibitemShut {NoStop}%
%%CITATION = ARXIV:1606.07609;%%
\bibitem [{\citenamefont {Jacques}\ \emph {et~al.}(2016)\citenamefont
  {Jacques}, \citenamefont {Katz}, \citenamefont {Morgante}, \citenamefont
  {Racco}, \citenamefont {Rameez},\ and\ \citenamefont
  {Riotto}}]{Jacques:2016dqz}%
  \BibitemOpen
  \bibfield  {author} {\bibinfo {author} {\bibfnamefont {T.}~\bibnamefont
  {Jacques}}, \bibinfo {author} {\bibfnamefont {A.}~\bibnamefont {Katz}},
  \bibinfo {author} {\bibfnamefont {E.}~\bibnamefont {Morgante}}, \bibinfo
  {author} {\bibfnamefont {D.}~\bibnamefont {Racco}}, \bibinfo {author}
  {\bibfnamefont {M.}~\bibnamefont {Rameez}}, \ and\ \bibinfo {author}
  {\bibfnamefont {A.}~\bibnamefont {Riotto}},\ }\href {\doibase
  10.1007/JHEP10(2016)071} {\bibfield  {journal} {\bibinfo  {journal} {JHEP}\
  }\textbf {\bibinfo {volume} {10}},\ \bibinfo {pages} {071} (\bibinfo {year}
  {2016})},\ \Eprint {http://arxiv.org/abs/1605.06513} {arXiv:1605.06513
  [hep-ph]} \BibitemShut {NoStop}%
%%CITATION = ARXIV:1605.06513;%%
\bibitem [{\citenamefont {Bell}\ \emph
  {et~al.}(2017{\natexlab{b}})\citenamefont {Bell}, \citenamefont {Cai},\ and\
  \citenamefont {Leane}}]{Bell:2016uhg}%
  \BibitemOpen
  \bibfield  {author} {\bibinfo {author} {\bibfnamefont {N.~F.}\ \bibnamefont
  {Bell}}, \bibinfo {author} {\bibfnamefont {Y.}~\bibnamefont {Cai}}, \ and\
  \bibinfo {author} {\bibfnamefont {R.~K.}\ \bibnamefont {Leane}},\ }\href
  {\doibase 10.1088/1475-7516/2017/01/039} {\bibfield  {journal} {\bibinfo
  {journal} {JCAP}\ }\textbf {\bibinfo {volume} {1701}},\ \bibinfo {pages}
  {039} (\bibinfo {year} {2017}{\natexlab{b}})},\ \Eprint
  {http://arxiv.org/abs/1610.03063} {arXiv:1610.03063 [hep-ph]} \BibitemShut
  {NoStop}%
%%CITATION = ARXIV:1610.03063;%%
\bibitem [{\citenamefont {D'Eramo}\ \emph {et~al.}(2016)\citenamefont
  {D'Eramo}, \citenamefont {Kavanagh},\ and\ \citenamefont
  {Panci}}]{DEramo:2016gos}%
  \BibitemOpen
  \bibfield  {author} {\bibinfo {author} {\bibfnamefont {F.}~\bibnamefont
  {D'Eramo}}, \bibinfo {author} {\bibfnamefont {B.~J.}\ \bibnamefont
  {Kavanagh}}, \ and\ \bibinfo {author} {\bibfnamefont {P.}~\bibnamefont
  {Panci}},\ }\href {\doibase 10.1007/JHEP08(2016)111} {\bibfield  {journal}
  {\bibinfo  {journal} {JHEP}\ }\textbf {\bibinfo {volume} {08}},\ \bibinfo
  {pages} {111} (\bibinfo {year} {2016})},\ \Eprint
  {http://arxiv.org/abs/1605.04917} {arXiv:1605.04917 [hep-ph]} \BibitemShut
  {NoStop}%
%%CITATION = ARXIV:1605.04917;%%
\bibitem [{\citenamefont {Englert}\ \emph {et~al.}(2016)\citenamefont
  {Englert}, \citenamefont {McCullough},\ and\ \citenamefont
  {Spannowsky}}]{Englert:2016joy}%
  \BibitemOpen
  \bibfield  {author} {\bibinfo {author} {\bibfnamefont {C.}~\bibnamefont
  {Englert}}, \bibinfo {author} {\bibfnamefont {M.}~\bibnamefont {McCullough}},
  \ and\ \bibinfo {author} {\bibfnamefont {M.}~\bibnamefont {Spannowsky}},\
  }\href {\doibase 10.1016/j.dark.2016.09.002} {\bibfield  {journal} {\bibinfo
  {journal} {Phys. Dark Univ.}\ }\textbf {\bibinfo {volume} {14}},\ \bibinfo
  {pages} {48} (\bibinfo {year} {2016})},\ \Eprint
  {http://arxiv.org/abs/1604.07975} {arXiv:1604.07975 [hep-ph]} \BibitemShut
  {NoStop}%
%%CITATION = ARXIV:1604.07975;%%
\bibitem [{\citenamefont {Brennan}\ \emph {et~al.}(2016)\citenamefont
  {Brennan}, \citenamefont {McDonald}, \citenamefont {Gramling},\ and\
  \citenamefont {Jacques}}]{Brennan:2016xjh}%
  \BibitemOpen
  \bibfield  {author} {\bibinfo {author} {\bibfnamefont {A.~J.}\ \bibnamefont
  {Brennan}}, \bibinfo {author} {\bibfnamefont {M.~F.}\ \bibnamefont
  {McDonald}}, \bibinfo {author} {\bibfnamefont {J.}~\bibnamefont {Gramling}},
  \ and\ \bibinfo {author} {\bibfnamefont {T.~D.}\ \bibnamefont {Jacques}},\
  }\href {\doibase 10.1007/JHEP05(2016)112} {\bibfield  {journal} {\bibinfo
  {journal} {JHEP}\ }\textbf {\bibinfo {volume} {05}},\ \bibinfo {pages} {112}
  (\bibinfo {year} {2016})},\ \Eprint {http://arxiv.org/abs/1603.01366}
  {arXiv:1603.01366 [hep-ph]} \BibitemShut {NoStop}%
%%CITATION = ARXIV:1603.01366;%%
\bibitem [{\citenamefont {De~Romeri}\ \emph {et~al.}(2017)\citenamefont
  {De~Romeri}, \citenamefont {Fernandez-Martinez}, \citenamefont {Gehrlein},
  \citenamefont {Machado},\ and\ \citenamefont {Niro}}]{DeRomeri:2017oxa}%
  \BibitemOpen
  \bibfield  {author} {\bibinfo {author} {\bibfnamefont {V.}~\bibnamefont
  {De~Romeri}}, \bibinfo {author} {\bibfnamefont {E.}~\bibnamefont
  {Fernandez-Martinez}}, \bibinfo {author} {\bibfnamefont {J.}~\bibnamefont
  {Gehrlein}}, \bibinfo {author} {\bibfnamefont {P.~A.~N.}\ \bibnamefont
  {Machado}}, \ and\ \bibinfo {author} {\bibfnamefont {V.}~\bibnamefont
  {Niro}},\ }\href {\doibase 10.1007/JHEP10(2017)169} {\bibfield  {journal}
  {\bibinfo  {journal} {JHEP}\ }\textbf {\bibinfo {volume} {10}},\ \bibinfo
  {pages} {169} (\bibinfo {year} {2017})},\ \Eprint
  {http://arxiv.org/abs/1707.08606} {arXiv:1707.08606 [hep-ph]} \BibitemShut
  {NoStop}%
%%CITATION = ARXIV:1707.08606;%%
\bibitem [{\citenamefont {Cui}\ and\ \citenamefont
  {D'Eramo}(2017)}]{Cui:2017juz}%
  \BibitemOpen
  \bibfield  {author} {\bibinfo {author} {\bibfnamefont {Y.}~\bibnamefont
  {Cui}}\ and\ \bibinfo {author} {\bibfnamefont {F.}~\bibnamefont {D'Eramo}},\
  }\href {\doibase 10.1103/PhysRevD.96.095006} {\bibfield  {journal} {\bibinfo
  {journal} {Phys. Rev.}\ }\textbf {\bibinfo {volume} {D96}},\ \bibinfo {pages}
  {095006} (\bibinfo {year} {2017})},\ \Eprint
  {http://arxiv.org/abs/1705.03897} {arXiv:1705.03897 [hep-ph]} \BibitemShut
  {NoStop}%
%%CITATION = ARXIV:1705.03897;%%
\bibitem [{\citenamefont {Mantilla}\ and\ \citenamefont
  {Martinez}(2017)}]{Mantilla:2017ijh}%
  \BibitemOpen
  \bibfield  {author} {\bibinfo {author} {\bibfnamefont {S.~F.}\ \bibnamefont
  {Mantilla}}\ and\ \bibinfo {author} {\bibfnamefont {R.}~\bibnamefont
  {Martinez}},\ }\href {\doibase 10.1103/PhysRevD.96.095027} {\bibfield
  {journal} {\bibinfo  {journal} {Phys. Rev.}\ }\textbf {\bibinfo {volume}
  {D96}},\ \bibinfo {pages} {095027} (\bibinfo {year} {2017})},\ \Eprint
  {http://arxiv.org/abs/1704.04869} {arXiv:1704.04869 [hep-ph]} \BibitemShut
  {NoStop}%
%%CITATION = ARXIV:1704.04869;%%
\bibitem [{\citenamefont {Bauer}\ \emph {et~al.}(2018)\citenamefont {Bauer},
  \citenamefont {Diefenbacher}, \citenamefont {Plehn}, \citenamefont
  {Russell},\ and\ \citenamefont {Camargo}}]{Bauer:2018egk}%
  \BibitemOpen
  \bibfield  {author} {\bibinfo {author} {\bibfnamefont {M.}~\bibnamefont
  {Bauer}}, \bibinfo {author} {\bibfnamefont {S.}~\bibnamefont {Diefenbacher}},
  \bibinfo {author} {\bibfnamefont {T.}~\bibnamefont {Plehn}}, \bibinfo
  {author} {\bibfnamefont {M.}~\bibnamefont {Russell}}, \ and\ \bibinfo
  {author} {\bibfnamefont {D.~A.}\ \bibnamefont {Camargo}},\ }\href@noop {} {\
  (\bibinfo {year} {2018})},\ \Eprint {http://arxiv.org/abs/1805.01904}
  {arXiv:1805.01904 [hep-ph]} \BibitemShut {NoStop}%
%%CITATION = ARXIV:1805.01904;%%
\bibitem [{\citenamefont {Aprile}\ \emph {et~al.}(2018)\citenamefont {Aprile}
  \emph {et~al.}}]{Aprile:2018dbl}%
  \BibitemOpen
  \bibfield  {author} {\bibinfo {author} {\bibfnamefont {E.}~\bibnamefont
  {Aprile}} \emph {et~al.} (\bibinfo {collaboration} {XENON}),\ }\href@noop {}
  {\  (\bibinfo {year} {2018})},\ \Eprint {http://arxiv.org/abs/1805.12562}
  {arXiv:1805.12562 [astro-ph.CO]} \BibitemShut {NoStop}%
%%CITATION = ARXIV:1805.12562;%%
\bibitem [{\citenamefont {Akerib}\ \emph {et~al.}(2016)\citenamefont {Akerib}
  \emph {et~al.}}]{Akerib:2016vxi}%
  \BibitemOpen
  \bibfield  {author} {\bibinfo {author} {\bibfnamefont {D.~S.}\ \bibnamefont
  {Akerib}} \emph {et~al.},\ }\href@noop {} {\  (\bibinfo {year} {2016})},\
  \Eprint {http://arxiv.org/abs/1608.07648} {arXiv:1608.07648 [astro-ph.CO]}
  \BibitemShut {NoStop}%
%%CITATION = ARXIV:1608.07648;%%
\bibitem [{\citenamefont {Tan}\ \emph {et~al.}(2016)\citenamefont {Tan} \emph
  {et~al.}}]{Tan:2016zwf}%
  \BibitemOpen
  \bibfield  {author} {\bibinfo {author} {\bibfnamefont {A.}~\bibnamefont
  {Tan}} \emph {et~al.} (\bibinfo {collaboration} {PandaX-II}),\ }\href
  {\doibase 10.1103/PhysRevLett.117.121303} {\bibfield  {journal} {\bibinfo
  {journal} {Phys. Rev. Lett.}\ }\textbf {\bibinfo {volume} {117}},\ \bibinfo
  {pages} {121303} (\bibinfo {year} {2016})},\ \Eprint
  {http://arxiv.org/abs/1607.07400} {arXiv:1607.07400 [hep-ex]} \BibitemShut
  {NoStop}%
%%CITATION = ARXIV:1607.07400;%%
\bibitem [{\citenamefont {Cui}\ \emph {et~al.}(2017)\citenamefont {Cui} \emph
  {et~al.}}]{Cui:2017nnn}%
  \BibitemOpen
  \bibfield  {author} {\bibinfo {author} {\bibfnamefont {X.}~\bibnamefont
  {Cui}} \emph {et~al.} (\bibinfo {collaboration} {PandaX-II}),\ }\href
  {\doibase 10.1103/PhysRevLett.119.181302} {\bibfield  {journal} {\bibinfo
  {journal} {Phys. Rev. Lett.}\ }\textbf {\bibinfo {volume} {119}},\ \bibinfo
  {pages} {181302} (\bibinfo {year} {2017})},\ \Eprint
  {http://arxiv.org/abs/1708.06917} {arXiv:1708.06917 [astro-ph.CO]}
  \BibitemShut {NoStop}%
%%CITATION = ARXIV:1708.06917;%%
\bibitem [{\citenamefont {Belanger}\ \emph {et~al.}(2007)\citenamefont
  {Belanger}, \citenamefont {Boudjema}, \citenamefont {Pukhov},\ and\
  \citenamefont {Semenov}}]{Belanger:2006is}%
  \BibitemOpen
  \bibfield  {author} {\bibinfo {author} {\bibfnamefont {G.}~\bibnamefont
  {Belanger}}, \bibinfo {author} {\bibfnamefont {F.}~\bibnamefont {Boudjema}},
  \bibinfo {author} {\bibfnamefont {A.}~\bibnamefont {Pukhov}}, \ and\ \bibinfo
  {author} {\bibfnamefont {A.}~\bibnamefont {Semenov}},\ }\href {\doibase
  10.1016/j.cpc.2006.11.008} {\bibfield  {journal} {\bibinfo  {journal}
  {Comput. Phys. Commun.}\ }\textbf {\bibinfo {volume} {176}},\ \bibinfo
  {pages} {367} (\bibinfo {year} {2007})},\ \Eprint
  {http://arxiv.org/abs/hep-ph/0607059} {arXiv:hep-ph/0607059 [hep-ph]}
  \BibitemShut {NoStop}%
%%CITATION = HEP-PH/0607059;%%
\bibitem [{\citenamefont {Belanger}\ \emph {et~al.}(2009)\citenamefont
  {Belanger}, \citenamefont {Boudjema}, \citenamefont {Pukhov},\ and\
  \citenamefont {Semenov}}]{Belanger:2008sj}%
  \BibitemOpen
  \bibfield  {author} {\bibinfo {author} {\bibfnamefont {G.}~\bibnamefont
  {Belanger}}, \bibinfo {author} {\bibfnamefont {F.}~\bibnamefont {Boudjema}},
  \bibinfo {author} {\bibfnamefont {A.}~\bibnamefont {Pukhov}}, \ and\ \bibinfo
  {author} {\bibfnamefont {A.}~\bibnamefont {Semenov}},\ }\href {\doibase
  10.1016/j.cpc.2008.11.019} {\bibfield  {journal} {\bibinfo  {journal}
  {Comput. Phys. Commun.}\ }\textbf {\bibinfo {volume} {180}},\ \bibinfo
  {pages} {747} (\bibinfo {year} {2009})},\ \Eprint
  {http://arxiv.org/abs/0803.2360} {arXiv:0803.2360 [hep-ph]} \BibitemShut
  {NoStop}%
%%CITATION = ARXIV:0803.2360;%%
\bibitem [{\citenamefont {Belanger}\ \emph {et~al.}(2014)\citenamefont
  {Belanger}, \citenamefont {Boudjema}, \citenamefont {Pukhov},\ and\
  \citenamefont {Semenov}}]{Belanger:2013oya}%
  \BibitemOpen
  \bibfield  {author} {\bibinfo {author} {\bibfnamefont {G.}~\bibnamefont
  {Belanger}}, \bibinfo {author} {\bibfnamefont {F.}~\bibnamefont {Boudjema}},
  \bibinfo {author} {\bibfnamefont {A.}~\bibnamefont {Pukhov}}, \ and\ \bibinfo
  {author} {\bibfnamefont {A.}~\bibnamefont {Semenov}},\ }\href {\doibase
  10.1016/j.cpc.2013.10.016} {\bibfield  {journal} {\bibinfo  {journal}
  {Comput. Phys. Commun.}\ }\textbf {\bibinfo {volume} {185}},\ \bibinfo
  {pages} {960} (\bibinfo {year} {2014})},\ \Eprint
  {http://arxiv.org/abs/1305.0237} {arXiv:1305.0237 [hep-ph]} \BibitemShut
  {NoStop}%
%%CITATION = ARXIV:1305.0237;%%
\bibitem [{\citenamefont {Ade}\ \emph {et~al.}(2016)\citenamefont {Ade} \emph
  {et~al.}}]{Ade:2015xua}%
  \BibitemOpen
  \bibfield  {author} {\bibinfo {author} {\bibfnamefont {P.~A.~R.}\
  \bibnamefont {Ade}} \emph {et~al.} (\bibinfo {collaboration} {Planck}),\
  }\href {\doibase 10.1051/0004-6361/201525830} {\bibfield  {journal} {\bibinfo
   {journal} {Astron. Astrophys.}\ }\textbf {\bibinfo {volume} {594}},\
  \bibinfo {pages} {A13} (\bibinfo {year} {2016})},\ \Eprint
  {http://arxiv.org/abs/1502.01589} {arXiv:1502.01589 [astro-ph.CO]}
  \BibitemShut {NoStop}%
%%CITATION = ARXIV:1502.01589;%%
\bibitem [{\citenamefont {Sjostrand}\ \emph {et~al.}(2008)\citenamefont
  {Sjostrand}, \citenamefont {Mrenna},\ and\ \citenamefont
  {Skands}}]{Sjostrand:2007gs}%
  \BibitemOpen
  \bibfield  {author} {\bibinfo {author} {\bibfnamefont {T.}~\bibnamefont
  {Sjostrand}}, \bibinfo {author} {\bibfnamefont {S.}~\bibnamefont {Mrenna}}, \
  and\ \bibinfo {author} {\bibfnamefont {P.~Z.}\ \bibnamefont {Skands}},\
  }\href {\doibase 10.1016/j.cpc.2008.01.036} {\bibfield  {journal} {\bibinfo
  {journal} {Comput. Phys. Commun.}\ }\textbf {\bibinfo {volume} {178}},\
  \bibinfo {pages} {852} (\bibinfo {year} {2008})},\ \Eprint
  {http://arxiv.org/abs/0710.3820} {arXiv:0710.3820 [hep-ph]} \BibitemShut
  {NoStop}%
%%CITATION = ARXIV:0710.3820;%%
\bibitem [{\citenamefont {Kong}\ and\ \citenamefont
  {Park}(2014)}]{Kong:2014haa}%
  \BibitemOpen
  \bibfield  {author} {\bibinfo {author} {\bibfnamefont {K.}~\bibnamefont
  {Kong}}\ and\ \bibinfo {author} {\bibfnamefont {J.-C.}\ \bibnamefont
  {Park}},\ }\href {\doibase 10.1016/j.nuclphysb.2014.09.014} {\bibfield
  {journal} {\bibinfo  {journal} {Nucl. Phys.}\ }\textbf {\bibinfo {volume}
  {B888}},\ \bibinfo {pages} {154} (\bibinfo {year} {2014})},\ \Eprint
  {http://arxiv.org/abs/1404.3741} {arXiv:1404.3741 [hep-ph]} \BibitemShut
  {NoStop}%
%%CITATION = ARXIV:1404.3741;%%
\bibitem [{\citenamefont {Dutta}\ \emph {et~al.}(2015)\citenamefont {Dutta},
  \citenamefont {Gao}, \citenamefont {Ghosh},\ and\ \citenamefont
  {Strigari}}]{Dutta:2015ysa}%
  \BibitemOpen
  \bibfield  {author} {\bibinfo {author} {\bibfnamefont {B.}~\bibnamefont
  {Dutta}}, \bibinfo {author} {\bibfnamefont {Y.}~\bibnamefont {Gao}}, \bibinfo
  {author} {\bibfnamefont {T.}~\bibnamefont {Ghosh}}, \ and\ \bibinfo {author}
  {\bibfnamefont {L.~E.}\ \bibnamefont {Strigari}},\ }\href {\doibase
  10.1103/PhysRevD.92.075019} {\bibfield  {journal} {\bibinfo  {journal} {Phys.
  Rev.}\ }\textbf {\bibinfo {volume} {D92}},\ \bibinfo {pages} {075019}
  (\bibinfo {year} {2015})},\ \Eprint {http://arxiv.org/abs/1508.05989}
  {arXiv:1508.05989 [hep-ph]} \BibitemShut {NoStop}%
%%CITATION = ARXIV:1508.05989;%%
\bibitem [{\citenamefont {Duerr}\ \emph {et~al.}(2015)\citenamefont {Duerr},
  \citenamefont {Fileviez~Perez},\ and\ \citenamefont
  {Smirnov}}]{Duerr:2015wfa}%
  \BibitemOpen
  \bibfield  {author} {\bibinfo {author} {\bibfnamefont {M.}~\bibnamefont
  {Duerr}}, \bibinfo {author} {\bibfnamefont {P.}~\bibnamefont
  {Fileviez~Perez}}, \ and\ \bibinfo {author} {\bibfnamefont {J.}~\bibnamefont
  {Smirnov}},\ }\href@noop {} {\  (\bibinfo {year} {2015})},\ \Eprint
  {http://arxiv.org/abs/1506.05107} {arXiv:1506.05107 [hep-ph]} \BibitemShut
  {NoStop}%
%%CITATION = ARXIV:1506.05107;%%
\bibitem [{\citenamefont {Profumo}\ \emph {et~al.}(2016)\citenamefont
  {Profumo}, \citenamefont {Queiroz},\ and\ \citenamefont
  {Yaguna}}]{Profumo:2016idl}%
  \BibitemOpen
  \bibfield  {author} {\bibinfo {author} {\bibfnamefont {S.}~\bibnamefont
  {Profumo}}, \bibinfo {author} {\bibfnamefont {F.~S.}\ \bibnamefont
  {Queiroz}}, \ and\ \bibinfo {author} {\bibfnamefont {C.~E.}\ \bibnamefont
  {Yaguna}},\ }\href {\doibase 10.1093/mnras/stw1600} {\bibfield  {journal}
  {\bibinfo  {journal} {Mon. Not. Roy. Astron. Soc.}\ }\textbf {\bibinfo
  {volume} {461}},\ \bibinfo {pages} {3976} (\bibinfo {year} {2016})},\ \Eprint
  {http://arxiv.org/abs/1602.08501} {arXiv:1602.08501 [astro-ph.HE]}
  \BibitemShut {NoStop}%
%%CITATION = ARXIV:1602.08501;%%
\bibitem [{\citenamefont {Aaboud}\ \emph {et~al.}(2017)\citenamefont {Aaboud}
  \emph {et~al.}}]{Aaboud:2017buh}%
  \BibitemOpen
  \bibfield  {author} {\bibinfo {author} {\bibfnamefont {M.}~\bibnamefont
  {Aaboud}} \emph {et~al.} (\bibinfo {collaboration} {ATLAS}),\ }\href
  {\doibase 10.1007/JHEP10(2017)182} {\bibfield  {journal} {\bibinfo  {journal}
  {JHEP}\ }\textbf {\bibinfo {volume} {10}},\ \bibinfo {pages} {182} (\bibinfo
  {year} {2017})},\ \Eprint {http://arxiv.org/abs/1707.02424} {arXiv:1707.02424
  [hep-ex]} \BibitemShut {NoStop}%
%%CITATION = ARXIV:1707.02424;%%
\bibitem [{\citenamefont {Camargo}\ \emph {et~al.}(2018)\citenamefont
  {Camargo}, \citenamefont {Delle~Rose}, \citenamefont {Moretti},\ and\
  \citenamefont {Queiroz}}]{Camargo:2018klg}%
  \BibitemOpen
  \bibfield  {author} {\bibinfo {author} {\bibfnamefont {D.~A.}\ \bibnamefont
  {Camargo}}, \bibinfo {author} {\bibfnamefont {L.}~\bibnamefont {Delle~Rose}},
  \bibinfo {author} {\bibfnamefont {S.}~\bibnamefont {Moretti}}, \ and\
  \bibinfo {author} {\bibfnamefont {F.~S.}\ \bibnamefont {Queiroz}},\
  }\href@noop {} {\  (\bibinfo {year} {2018})},\ \Eprint
  {http://arxiv.org/abs/1805.08231} {arXiv:1805.08231 [hep-ph]} \BibitemShut
  {NoStop}%
%%CITATION = ARXIV:1805.08231;%%
\bibitem [{\citenamefont {Papucci}\ \emph {et~al.}(2014)\citenamefont
  {Papucci}, \citenamefont {Sakurai}, \citenamefont {Weiler},\ and\
  \citenamefont {Zeune}}]{Papucci:2014rja}%
  \BibitemOpen
  \bibfield  {author} {\bibinfo {author} {\bibfnamefont {M.}~\bibnamefont
  {Papucci}}, \bibinfo {author} {\bibfnamefont {K.}~\bibnamefont {Sakurai}},
  \bibinfo {author} {\bibfnamefont {A.}~\bibnamefont {Weiler}}, \ and\ \bibinfo
  {author} {\bibfnamefont {L.}~\bibnamefont {Zeune}},\ }\href {\doibase
  10.1140/epjc/s10052-014-3163-1} {\bibfield  {journal} {\bibinfo  {journal}
  {Eur. Phys. J.}\ }\textbf {\bibinfo {volume} {C74}},\ \bibinfo {pages} {3163}
  (\bibinfo {year} {2014})},\ \Eprint {http://arxiv.org/abs/1402.0492}
  {arXiv:1402.0492 [hep-ph]} \BibitemShut {NoStop}%
%%CITATION = ARXIV:1402.0492;%%
\end{thebibliography}%

\end{document}